%% file: main.tex
\documentclass[11pt]{article}

\usepackage[T1]{fontenc}
\usepackage[utf8]{inputenc}
\usepackage[margin=1in]{geometry}
\usepackage{amsmath,amssymb,amsthm}
\usepackage{aliascnt}
\usepackage{mathtools}
\usepackage{booktabs}
\usepackage{array}
\usepackage{tikz}
\newcolumntype{P}[1]{>{\raggedright\arraybackslash}p{#1}}
\usepackage{microtype}
\usepackage{xcolor}
\usepackage{cite}
\usepackage[colorlinks=true,linkcolor=blue,citecolor=blue,urlcolor=blue]{hyperref}
\usepackage[nameinlink,noabbrev,capitalise]{cleveref}

\newtheorem{theorem}{Theorem}[section]
\newaliascnt{lemma}{theorem}
\newtheorem{lemma}[lemma]{Lemma}
\aliascntresetthe{lemma}
\newaliascnt{proposition}{theorem}
\newtheorem{proposition}[proposition]{Proposition}
\aliascntresetthe{proposition}
\newaliascnt{corollary}{theorem}
\newtheorem{corollary}[corollary]{Corollary}
\aliascntresetthe{corollary}
\theoremstyle{definition}
\newaliascnt{definition}{theorem}
\newtheorem{definition}[definition]{Definition}
\aliascntresetthe{definition}
\newaliascnt{example}{theorem}
\newtheorem{example}[example]{Example}
\aliascntresetthe{example}
\theoremstyle{remark}
\newaliascnt{remark}{theorem}
\newtheorem{remark}[remark]{Remark}
\aliascntresetthe{remark}

\newcommand{\R}{\mathbb{R}}
\newcommand{\hlst}{\mathrm{HL}^{\ast}}
\newcommand{\hlts}{\mathrm{HL}^{\mathrm{ts}}}
\newcommand{\hl}{\mathrm{HL}}
\newcommand{\Lz}{\Lambda_{0}}

\title{The Value of Steiner Hubs in Two-Hop Distance Labeling}
\author{Sichen Wang\\
Shenzhen MSU-BIT University\\
\texttt{wsc@smbu.edu.cn}}
\date{}

\begin{document}
\maketitle

\begin{abstract}
Two-hop distance labelings were introduced in 2002 by Cohen, Halperin, Kaplan, and Zwick
with hubs of two kinds: vertices of the graph, or, in their Steiner variant, arbitrary
objects carrying arbitrary real values. On no family is the value of that freedom known.
We determine it on $\ell_1$ products of betweenness-free metrics, those with no point
between two others, a class containing every Hamming graph. The freedom is worth nothing
when the factors have at most three letters, while from four letters on the largest saving
is polynomial in the size of the product, attained on products of positively weighted
stars. The $d$-dimensional hypercube is the two-letter case, and labeling it takes
$(5/2+o(1))^d$ entries even with arbitrary hubs, matching the vertex-hub optimum of
Goldberg, Razenshteyn, and Savchenko. Every lower bound here comes from one semidefinite
program on a single factor, whose value $\Lambda$ never exceeds the rate that vertex hubs
already achieve on the Hamming graph; where $\Lambda$ meets that ceiling, the freedom
is worth nothing, and the ceiling is met exactly up to three letters.
\end{abstract}

\input{sections/intro}

\input{sections/prelim}

\input{sections/invariant}

\input{sections/hamming}

\input{sections/grids}

\clearpage

\input{references}
\end{document}

%% file: sections/intro.tex
\section{Introduction}\label{sec:intro}

Whether auxiliary points help is a question that metric algorithms answer one structure
at a time, and the answers cover every scale. In a tree metric, Steiner points do not
really help: a tree on the terminals alone preserves their distances within a factor of
eight, and eight is tight~\cite{Gup01,CXKR06}; on general graphs removing them costs
distortion $O(\log k)$ and at least $\tilde\Omega(\sqrt{\log k})$ for $k$
terminals~\cite{Fil19,CT24a}. They make shallow-light trees exponentially
lighter~\cite{ES11}, they improve the size of Euclidean $(1+\varepsilon)$-spanners
quadratically~\cite{LS19,BT21}, they do not close the integrality gap of
$0$-extension~\cite{CT24b}, and they cut the number of DAGs that cover the distances of a
bounded-treewidth digraph exactly, at near-linear cost, from $\Omega(\log n)$ to
two~\cite{BCCFOWZ26}. For the exact two-hop representation of a metric the question is as
old as the representation, and it is the subject of this paper.

A two-hop distance label stores at each point a short list of hubs, each with a number,
and answers a query on two points from their two lists, as the smallest sum over the hubs
they share. No sum falls below the distance, a hub whose sum equals it is tight on the
pair, and exactness means that every pair has a tight hub. Cohen, Halperin, Kaplan, and
Zwick introduced the format in two
forms~\cite{CHKZ02}: hubs may be vertices of the graph, or, in their Definition~3.3,
elements of an arbitrary finite set carrying arbitrary real values, which they call a
Steiner labeling. A vertex is a special hub, so the second form contains the first, and
both measure size by the number of stored entries. The first form is the hub labeling
studied since, with an approximation and hardness theory~\cite{BGGN16,AMO17}, exact optima
on the hypercube~\cite{GRS13}, structural parameters that explain when labels are
small~\cite{ADFGW16,KV17}, and indexes at scale~\cite{AIY13,DGSW14}. The same paper left
a forecast for each form. On the vertex side, Conjecture~5.1 asks for two-hop
covers of all shortest paths of total size $O(n\sqrt m)$ in every directed graph. On the
Steiner side, Corollary~5.3 states that hubs off the graph save a factor of $\sqrt m$ on a
family of undirected graphs, with the proof deferred to a construction similar to the one
given for reachability.

Two decades on, the forecasts have fared differently. The vertex branch was carried to
completion: an $O(\log n)$ approximation for every $\ell_p$ objective~\cite{BGGN16},
NP-hardness~\cite{Wel14,Whi15,BGKSW15}, a matching $\Omega(\log n)$ inapproximability with
a PTAS on trees~\cite{AMO17}, and, on sparse graphs, the lower bound of Kosowski,
Uzna\'nski, and Viennot, who exhibit graphs of maximum degree three on which every hub
labeling has total size at least $n^{2}/2^{O(\sqrt{\log n})}$~\cite{KUV19}. A cover of all shortest
paths is in particular a hub labeling, so their bound also settles Conjecture~5.1, in the
negative.\footnote{The conjecture is stated for directed graphs. Orient each edge of such a
graph both ways, which doubles $m$, and merge the two lists of every vertex; a two-hop cover
of all shortest paths becomes a hub labeling of no larger total size, which
\cite[Thm.~1.1]{KUV19} makes at least $n^{2}/2^{\Theta(\sqrt{\log n})}$, against the
conjectured $O(n\sqrt m)=O(n^{3/2})$.} The Steiner branch stayed where its founders left
it, for a mechanical reason: greedy set cover, with which Cohen et al.\ approximate the vertex form
within $O(\log n)$, needs a finite ground set of candidate hubs, and Definition~3.3
supplies none.

Lower bounds tell the same story. Those known for the format pin hubs to the points of the
space: on the path~\cite{AS87,BKKLLPT25}, on the cube, where the hypothesis says
so~\cite{HIS13}, and on sparse graphs~\cite{KUV19}. Two exceptions allow Steiner points,
both recent and both counting the edges of a two-hop spanner: on the line, for unit-spaced
points~\cite{LMS23}, and, in August 2026, on the cube, where Krauthgamer and Petruschka proved
that a two-hop $t$-spanner needs $(2^d)^{1+\Omega(1/t)}$ edges even with Steiner
vertices~\cite{KP26}, noting that the argument ``does not extend beyond two hops,'' since
three hops admit a constant-stretch spanner of near-linear size. Neither concerns the model
of Definition~3.3, where a hub carries two unrelated values at a point, and on no family is
the value of the freedom known.

\emph{What is the freedom worth, and can it be priced exactly?} We answer on the $\ell_1$
products of betweenness-free metrics: distances add over coordinates, and no point of a
factor lies between two others, in Menger's sense; every Hamming graph and the hypercube
are such products. Whether the freedom is worth anything there depends on the alphabet
alone, and the answer switches at four letters. The price comes from a single-letter
invariant that bounds the rate of every $\ell_1$ power from below, and the invariant is
pinned on both sides, from above by the very hubs it competes against.

\subsection{Our Results}\label{sec:results}

Write $\hlst(X)$ for the least total number of entries in an exact two-hop labeling of a
finite metric space $X$ with arbitrary hubs, $\hl(X)$ for the least with hubs at points of
$X$, called proper, and $\hlts(X)$ for the least with hubs at points of the tight span of
$X$, the injective hull of Isbell, where a hub is one function $f$ satisfying
$f(x)+f(y)\ge\delta(x,y)$ for all $x,y$. The three are nested, $\hlst\le\hlts\le\hl$, and
\Cref{sec:prelim} defines them.

Two extremes fix the range. On the uniform metric, where all distances are one, a single
hub at distance one half from every point is tight on every pair, so $\hlst=\hlts=2n$, the
floor that the two sides of a query impose on any labeling, while proper hubs need
$\Theta(n^2)$ entries (\Cref{prop:uniform}). The freedom is worth a factor of $n$ there.
On the hypercube it is worth nothing.

\begin{theorem}[the cube; \Cref{cor:cube}]\label{thm:intro-cube}
For the $d$-dimensional hypercube $Q_d$ and every $d\ge1$,
$\hlst(Q_d)\ge2\bigl((5/2)^d-(3/2)^d\bigr)$. Hence $\hlst(Q_d)=(5/2+o(1))^d$, and all three
sizes of the hypercube have exponential rate $5/2$, in that their $d$-th roots tend to
$5/2$.
\end{theorem}

Goldberg, Razenshteyn, and Savchenko determined the proper-hub size of the cube,
$(2.5+o(1))^d$~\cite{GRS13}. \Cref{thm:intro-cube} says that their construction is optimal
against hubs that are arbitrary objects carrying arbitrary real values, and that their
constant is the two-letter value of a law stated next. At stretch one, the bound
of~\cite{KP26} applies to the symmetric case of the same quantity and gives
$(2^d)^{1+\Omega(1)}$; here the exponent is exact, $\log_2(5/2)=1.3219\ldots$, and it is
the exponent of the vertex hubs.

On products of betweenness-free factors the answer is a law in the alphabet size $q$. Its
constants are $L_q=q(2q-3)/(q-1)$ for $q\ge3$, with $L_2=5/2$, and $\mu_q=q(q+3)/4$ for
$q\le5$ and $\mu_q=q\sqrt{q-1}$ for $q\ge5$, together with a trace term $R_q<L_q$ that
matters only at finite $d$ (\Cref{eq:constants}). The first two coincide for $q\le3$ and
separate from $q=4$ on, where $L_4=20/3$ and $\mu_4=7$.

\begin{theorem}[the dichotomy; \Cref{thm:dichotomy,thm:starrate,thm:unique}]
\label{thm:intro-dichotomy}
Fix $q\ge2$ and let $D$ range over $\ell_1$ products of $d$ betweenness-free metrics on
$q$ points each.
\begin{enumerate}
\item Every such $D$ has $\hlst(D)\ge2(L_q^{\,d}-R_q^{\,d})$, and its proper-hub covering
problem is that of the Hamming graph $H(d,q)$.
\item For $q\le3$ the three sizes of every such $D$ share the exponential rate $L_q=\mu_q$.
\item For $q\ge4$ the ratio $\hl(D)/\hlts(D)$ has exponential rate at most $\mu_q/L_q>1$,
with equality on products of positively weighted stars, and the stars are the only bases
on which the rate of $\hlst$ is $L_q$.
\end{enumerate}
\end{theorem}

Whether the freedom is worth anything on this cone is decided by the alphabet alone, and
so is its largest value: the lengths of the factors enter neither bound. The lower bound
holds against every labeling
in the general model, the upper bound uses hubs that are points of the tight span, one added
symbol per coordinate, and the unit is the same on both sides. On the cube the two rates
coincide, which is \Cref{thm:intro-cube}; at four letters they part, and from then on the
gap is polynomial in $n$. Letting the alphabet grow with the dimension turns the
dichotomy into a separation on one family: with $q_d=\max\{5,\lceil\ln d\rceil\}$ and
$n=q_d^{\,d}$, the Hamming graphs $H(d,q_d)$ have tight-span labelings of size
$n^{1+o(1)}$ while every proper labeling has size at least $n^{3/2-o(1)}$, on
distance-transitive
graphs of degree $(\ln n)^{1+o(1)}$ (\Cref{cor:family}).

Both theorems come from one invariant, and the invariant has a range.

\begin{theorem}[the invariant and its range; \Cref{prop:rate,thm:floor,thm:ceiling}]
\label{thm:intro-invariant}
Every metric space $X$ on $q\ge2$ points carries an invariant $\Lambda(X)$, the value of a
semidefinite program over nonnegative kernels on $X$ whose norm, masked by any set on which
one hub can be tight, is at most one. It satisfies
$\liminf_{d\to\infty}\hlst(X^{\oplus d})^{1/d}\ge\Lambda(X)$, and for every such $X$,
\[
  q\ <\ \Lz(q)\ \le\ \Lambda(X)\ \le\ \mu_q ,
  \qquad
  \Lz(q)=\frac{q\bigl(2q-1+2\sqrt{q-1}\bigr)}{q+2\sqrt{q-1}} ,
\]
and the ceiling $\mu_q$ is the proper-hub rate of $H(d,q)$. It is attained by every $X$ on
two points, by exactly the betweenness-free $X$ on three points, and by no $X$ on four or
more.
\end{theorem}

\begin{figure}[t]
\centering
\begin{tikzpicture}[x=0.72cm,y=13cm]
\fill[blue!8]
  (0,0.1719) -- (1,0.1185) -- (2,0.0937) -- (3,0.0785) -- (4,0.0678) -- (5,0.0598)
  -- (6,0.0535) -- (7,0.0482) -- (8,0.0438) -- (9,0.0400) -- (10,0.0367)
  -- (10,0.3325) -- (9,0.3301) -- (8,0.3271) -- (7,0.3232) -- (6,0.3179) -- (5,0.3104)
  -- (4,0.2991) -- (3,0.2807) -- (2,0.2537) -- (1,0.2191) -- cycle;
\draw[->] (0,0) -- (10.7,0) node[right] {$q$};
\draw[->] (0,0) -- (0,0.375) node[above] {\small exponent of $n$};
\foreach \q/\x in {2/0,3/1,4/2,5/3,6/4,7/5,8/6,9/7,10/8,11/9,12/10}
  {\draw (\x,0) -- (\x,-0.006) node[below,font=\scriptsize] {\q};}
\foreach \e/\y in {1.2/0.05,1.3/0.15,1.4/0.25,1.5/0.35}
  {\draw (0,\y) -- (-0.15,\y) node[left,font=\scriptsize] {\e};}
\draw[dashed,gray] (0,0.35) -- (10.4,0.35);
\draw[thick] plot[mark=*,mark size=1pt] coordinates
  {(0,0.1719) (1,0.2191) (2,0.2537) (3,0.2807) (4,0.2991) (5,0.3104) (6,0.3179)
   (7,0.3232) (8,0.3271) (9,0.3301) (10,0.3325)}
  node[right,font=\small] {$\log_q\mu_q$};
\draw[thick] plot[mark=square*,mark size=1pt] coordinates
  {(0,0.1719) (1,0.2191) (2,0.2185) (3,0.1977) (4,0.1780) (5,0.1615) (6,0.1477)
   (7,0.1361) (8,0.1262) (9,0.1177) (10,0.1102)}
  node[right,font=\small] {$\log_qL_q$};
\draw[thick] plot[mark=triangle*,mark size=1.3pt] coordinates
  {(0,0.1719) (1,0.1185) (2,0.0937) (3,0.0785) (4,0.0678) (5,0.0598) (6,0.0535)
   (7,0.0482) (8,0.0438) (9,0.0400) (10,0.0367)}
  node[right,font=\small] {$\log_q\Lz(q)$};
\end{tikzpicture}
\caption{The range of \Cref{thm:intro-invariant}, drawn as exponents of $n=q^d$: a rate
$r$ appears as $\log_qr$. The three curves coincide at $q=2$; $L_q$ and $\mu_q$ coincide
at $q=3$ and part from $q=4$, the gap between them being the value of Steiner freedom on
star products; $\Lz(q)$ runs below $L_q$ from $q=3$ on. The shaded band is where
$\Lambda(X)$ lies for every $X$ on $q$ points, and the dashed line is the exponent
$3/2$ that the proper-hub curve approaches.}
\label{fig:corridor}
\end{figure}
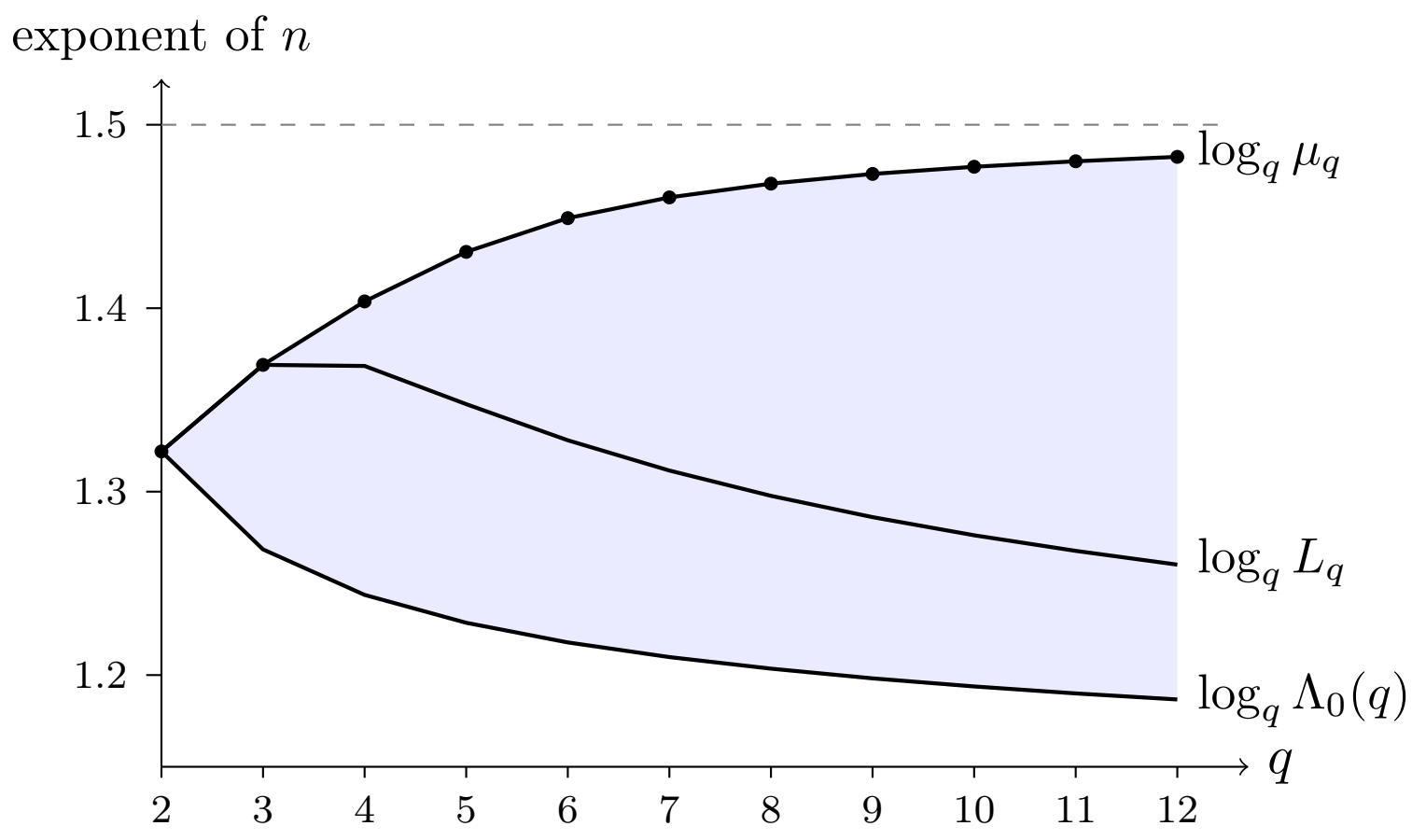

Two consequences read off directly. No $\ell_1$ power is free: on $n=q^d$ points,
$\hlst(X^{\oplus d})\ge n^{1+c_q-o(1)}$ with $c_q=\log_q(\Lz(q)/q)>0$, whatever the base.
And on a betweenness-free base the freedom is worth at most
$(\mu_q/\Lambda(X))^{d(1+o(1))}$, since proper hubs achieve $\mu_q$ there. So the method
stops exactly where proper hubs stand: it never proves a Steiner rate, the exponential rate
of $\hlst$, above the proper-hub rate; where $\Lambda(X)$ reaches $\mu_q$ the freedom is
worth nothing, and its shortfall below $\mu_q$ is the most the freedom can buy. On the
betweenness-free cone $\Lambda$ is smallest at $L_q$ (\Cref{thm:Lq}), so the shortfall is
largest there, and that is the dichotomy: $L_q=\mu_q$ up to three letters and $L_q<\mu_q$
from four. As
Lov\'asz's $\vartheta$ bounds the independence rate of strong powers from
above~\cite{Lov79}, $\Lambda$ bounds the Steiner rate of $\ell_1$ powers from below; it is
exactly multiplicative, and equal to the rate, on a class of bases closed under $\ell_1$
products that contains every positively weighted star and hence every Hamming graph
(\Cref{sec:exact}). \Cref{tab:landscape} collects the answer with its reasons, and
\Cref{fig:corridor} draws the range.

\begin{table}[t]
\centering\footnotesize
\begin{tabular}{@{}P{3.25cm}P{2.8cm}P{2.2cm}P{2.6cm}P{3.7cm}@{}}
\toprule
family & $\hlst$ & $\hl$ & $\hl/\hlst$ & where \\
\midrule
uniform metric on $n$ points & $2n$ & $\Theta(n^2)$ & $\Theta(n)$ &
  one hub at the center of the space (\Cref{prop:uniform}) \\
betweenness-free products, $q\le3$, the cube among them & $(L_q+o(1))^d$ & $(L_q+o(1))^d$ &
  $n^{o(1)}$ & the invariant meets its ceiling (\Cref{thm:dichotomy,cor:cube}) \\
betweenness-free products, $q\ge4$ & $\ge(L_q-o(1))^d$, and $(L_q+o(1))^d$ on star products &
  $(\mu_q+o(1))^d$ & $\le(\mu_q/L_q)^{d(1+o(1))}$, attained on star products &
  center coordinates at density $(q-3)/(2q-3)$ (\Cref{thm:dichotomy,thm:starrate}) \\
$H(d,q_d)$ with $q_d=\max\{5,\lceil\ln d\rceil\}$, $d$ large and even & $n^{1+o(1)}$ & $\ge n^{3/2-o(1)}$ &
  $\ge n^{1/2-o(1)}$ & bounds uniform in $q$ (\Cref{cor:family}) \\
grids $[a_1]\times\cdots\times[a_d]$, $d$ fixed & $\Theta_d(n\prod_i\log a_i)$ &
  $\Theta_d(n\prod_i\log a_i)$ & $O_d(1)$ &
  a line has no incidence off its points (\Cref{thm:grid,prop:vertexdom}) \\
\bottomrule
\end{tabular}
\caption{The value of Steiner freedom, with the mechanism that fixes it. Throughout,
$n$ is the number of points and $d$ the number of factors.}
\label{tab:landscape}
\end{table}

Conventions are those of \Cref{def:model}: size counts the entries stored for the two
sides of a query over all ordered pairs, so the one-sided count of~\cite{GRS13} differs
from ours by at most a factor of two. The labelings behind the upper bounds are explicit,
and those of \Cref{thm:starrate} are computable in polynomial time
(\Cref{rem:constructive}).

\subsection{Overview}\label{sec:overview}

Consider a lower bound. A labeling assigns each ordered pair to a hub tight on it, so the
question is what a single hub can be tight on, and the answer is a condition on the pairs
alone: a set $P$ of pairs lies in the tight set of some hub exactly when it is cyclically
monotone, that is, when no cyclic rematching of its pairs increases the total distance
(\Cref{prop:cyclic}). This is Rockafellar's theorem in its finite
form~\cite{Roc66,Roc87}, and membership is a negative-cycle test. Now weight the pairs by
a nonnegative kernel $K$ with zero diagonal and normalize it so that
$\|K\circ\mathbf 1_P\|\le1$ for every cyclically monotone $P$, the norm being spectral and
$\mathbf 1_P$ the indicator of $P$, its \emph{mask}. A hub with supports $A$ and $B$ then
collects weight at most $\sqrt{|A||B|}\le(|A|+|B|)/2$ from its tight set, so every
labeling has size at least $2\langle J,K\rangle$ (\Cref{lem:packing}). A matrix constrained
through a family of masked norms is the filtered $\gamma_2$ norm of~\cite{LMRSS11}; here
the metric dictates the filter.

Two moves turn this into exponential rates on products, and they are the paper's engine.
The first is tensorization at prefactor one: for kernels $K_1$ on $X_1$ and $K_2$ on $X_2$,
the mask norm of $K_1\otimes K_2$ on $X_1\oplus X_2$ is at most the product of the two mask
norms (\Cref{thm:tensor}). The tight set of a product hub is not a product of factor
tight sets, so the exact multiplicativity of $\gamma_2$-type norms under a fixed
filter~\cite{LSS08} does not apply; what separates is the tightness condition itself,
along a fibration that pairs the row's second coordinate with the column's first, and its
extremizers along each fiber are tight sets of genuine factor hubs. The second move is
diagonal completion. No labeling has to cover the diagonal, so \Cref{lem:packing} asks for
a kernel that vanishes there; but the constraint can be imposed after the product is formed
instead of before. A single-letter kernel may then carry diagonal mass, its tensor carries
that mass mostly off the diagonal, and deleting the diagonal of the tensor once costs only
the product of the traces (\Cref{lem:completion}).

The cube shows both moves at work. On two points at distance one the kernel
$\tfrac34I+\tfrac12(J-I)$ has mask norm exactly one: a cyclically monotone set contains at
most one diagonal cell, so a mask lies inside the off-diagonal, where the kernel has norm
$\tfrac12$, or inside the cross at a point, where it becomes
$\bigl(\begin{smallmatrix}3/4&1/2\\1/2&0\end{smallmatrix}\bigr)$, with eigenvalues $1$
and $-\tfrac14$. Its value is $\tfrac52$ and its trace $\tfrac32$, so
\[
  \hlst(Q_d)\ \ge\ 2\bigl[(\tfrac52)^d-(\tfrac32)^d\bigr],
\]
the bound of \Cref{thm:intro-cube}. A zero diagonal imposed letter by letter leaves
$\kappa(J-I)$ with $\kappa\le1$, of value at most $2$, and the bound $2\cdot2^d$, the floor
of \Cref{eq:floor} and nothing more. The kernels that do the work below differ in their
diagonals: zero on a line, $(q-2)/(q-1)$ per point on a betweenness-free base with
$q\ge3$, and $(q+\sqrt{q-1})/(q+2\sqrt{q-1})$ per point on an arbitrary base.

Optimizing the value over kernels of mask norm at most one gives $\Lambda(X)$
(\Cref{def:lambda}), and the lower-bound side of the paper is that program evaluated on
three kinds of base (\Cref{tab:kernels}). Read from above, the same program gives the
ceiling. The proper hub at a point $p$ is available in every metric, and its tight set
contains the cross at $p$, the pairs with $p$ as an end, so the cross is a mask; masking a
kernel by it leaves the diagonal entry at $p$ and
the rest of row $p$, and the norm constraint caps the squared $\ell_2$ norm of that row by
one minus the diagonal entry. Cauchy--Schwarz over the $q$ rows then caps
$\langle J,K\rangle$ at $\mu_q$ (\Cref{thm:ceiling}), and $\mu_q$ is exactly the rate at
which proper hubs label the Hamming graph (\Cref{thm:starrate}). Equality forces a single
kernel, and for $q\ge4$ one further mask, present in every metric, refutes it.

\begin{table}[t]
\centering\footnotesize
\setlength{\tabcolsep}{4.5pt}
\begin{tabular}{@{}llccl@{}}
\toprule
base & kernel & diagonal entry & value & proves \\
\midrule
two points & $\tfrac34I+\tfrac12(J-I)$ & $3/4$ & $5/2$ & \Cref{cor:cube} \\
betweenness-free, $q\ge3$ & $\bigl((q-2)I+(J-I)\bigr)/(q-1)$ & $(q-2)/(q-1)$ &
  $L_q$ & \Cref{thm:Lq,thm:unique} \\
arbitrary, $q$ points & $c\,(J-I)+t\,I$ & $t$ &
  $\Lz(q)$ & \Cref{thm:floor} \\
line on $a$ points & $|x-y|^{-1}/\bigl(\pi(1+3^{-1/2})\bigr)$ & $0$ &
  $\Theta(a\log a)$ & \Cref{thm:grid} \\
\bottomrule
\end{tabular}
\caption{One program, four kernels. Each kernel has mask norm at most one on its base,
so its value is a lower bound on $\Lambda$ there and, tensorized, on the Steiner rate of
every product of such bases. In the third row $c=1/(q-1+\sqrt{q-1})$ and
$t=(q+\sqrt{q-1})/(q+2\sqrt{q-1})$.}
\label{tab:kernels}
\end{table}

The upper side never leaves the alphabet. On a product of betweenness-free factors, a
proper hub $h$ is tight on $(x,y)$ exactly when each coordinate of $h$ equals the
corresponding coordinate of $x$ or of $y$ (\Cref{prop:oblivious}), so the proper covering
problem of every such product is the covering problem of $H(d,q)$, whatever the lengths.
When the factors are stars, one further hub per coordinate is available, the star's
center, a point of the tight span that is tight on a pair exactly when the pair differs in
that coordinate (\Cref{lem:span}). The hub set $\prod_i([q]\cup\{c_i\})$, an orbit program
under $S_q\wr S_d$ whose layers decompose into biregular blocks of closed-form density
(\Cref{lem:layer,lem:sandwich}), and greedy set cover give the rate $L_q$, at a hub whose
centers occupy a fraction $(q-3)/(2q-3)$ of the coordinates (\Cref{thm:starrate}). That
fraction is zero at $q=3$ and positive from $q=4$: below the threshold the optimal hub is a
point of the space, and above it a point of the span in a constant fraction of its
coordinates, which is the same transition that \Cref{thm:intro-invariant} describes as the
invariant falling short of its ceiling.

Each move rests on one property, and none of the three mentions the base, which is what
makes them transferable. Tensorization needs only that the tight sets of a product be
generated by fibering one factor and extremizing, which holds whenever the cost adds over
coordinates. Diagonal completion needs only a nonnegative kernel and a norm family that is
monotone under entrywise domination. The ceiling needs only a mask that the axioms force
into every instance; the cap it produces has the shape of the cap that certificate
complexity places on the positive-weight adversary bound in quantum query
complexity~\cite{SS06,Zha05}, and here the cap is a rate that proper hubs attain.

\Cref{sec:prelim} fixes the model, the composition of labelings across products, and the
characterization of tight sets. \Cref{sec:invariant} builds the invariant and proves
\Cref{thm:intro-invariant} (\Cref{thm:floor,thm:ceiling}). \Cref{sec:hamming} proves
\Cref{thm:intro-dichotomy} (\Cref{thm:dichotomy}) with \Cref{thm:intro-cube}
(\Cref{cor:cube}) and the family (\Cref{cor:family}). \Cref{sec:grids} prices the grids,
computes $\Lambda$ on three points, and closes with the class of bases on which $\Lambda$
is the rate.

\subsection{Related Work}\label{sec:related}

Four works stand nearest. Cohen et al.~\cite{CHKZ02} own the model, the unit, and the first
separation between its two forms, the factor $\sqrt m$ of their Corollary~5.3;
\Cref{prop:uniform} gives the exact values on the uniform metric. Goldberg, Razenshteyn,
and Savchenko~\cite{GRS13} computed the proper-hub rate of the cube by a linear program
over vertex hubs symmetrized under the automorphism group; \Cref{sec:midpoint} extends the
symmetrization to $q$ letters and to hubs off the space, and \Cref{cor:cube} shows that
their value is the Steiner value. Kosowski, Uzna\'nski, and Viennot~\cite{KUV19} proved the
strongest lower bound for proper hubs on sparse graphs, by a counting argument over
labelings whose hubs are vertices. Krauthgamer and Petruschka~\cite{KP26} proved
the first lower bound on the cube that allows Steiner vertices, $(2^d)^{1+\Omega(1/t)}$
edges for a two-hop $t$-spanner, by a reduction to locality-sensitive hashing; at $t=1$ it
is the symmetric case of \Cref{thm:intro-cube} with an unspecified exponent. Two
neighboring formats cannot pose the question at all: an exact hopset of diameter two is
this covering problem read as a graph augmentation, so its hubs are vertices by
construction~\cite{KoP22,BH23}, and distance labeling in bits leaves the decoder
unrestricted, so a bound on entries gives one on bits and not the
reverse~\cite{GPPR04,AGHP16,GKU16}.

The objective is old. The total support of a min-plus factorization is the weight of the
biclique cover it induces, the measure that is Lupanov's rectifier complexity at depth
two~\cite{Lup56,JS13} and that Chung, Erd\H{o}s, and Spencer introduced for
graphs~\cite{CES83,Tuz84}; here the rectangles are tight sets, and tropical ranks count
the factors instead~\cite{DSS05,CC12,Shi14}. Direct-product theorems
for covering measures are delicate: the direct-product question of Jukna and Sergeev for
rectifier complexity is settled only to within a logarithm~\cite{CILS17}, Boolean rank is
strictly submultiplicative already on $J-I$~\cite{HP22}, and the fractional binary rank is
not multiplicative~\cite{GK25}. On the $\Lambda$-exact class the Steiner rate multiplies
exactly (\Cref{prop:closure}). The invariant belongs to the single-letter tradition that
runs from Shannon's zero-error capacity through Lov\'asz's $\vartheta$ to asymptotic
spectra~\cite{Sha56,Lov79,KO98,Str88,Zui19,WZ26}; its architecture, a matrix constrained
through masked spectral norms, is the filtered $\gamma_2$ norm~\cite{LMRSS11}, and
$\gamma_2$-type norms are exactly multiplicative under tensoring~\cite{LSS08} when the
filter stays fixed, whereas here it changes with the metric. Spectral norms as lower
bounds for searching go back to Chazelle~\cite{Cha98}, harmonic weights for dominance to
Fredman~\cite{Fre81,PS18}, and $\gamma_2$ under Kronecker products gives the discrepancy
bounds for boxes of~\cite{MNT20}. The adversary bound of quantum query complexity has the
same constrained shape~\cite{HLS07}, and its positive-weight form is capped by certificate
complexity~\cite{SS06,Zha05}, as ours is capped by the proper-hub rate.

The masks are the cyclically monotone sets of optimal transport, in the maximizing
direction studied by Smith and Knott~\cite{SK92}; in mechanism design the same condition is
implementability~\cite{Roc87,SY05,ABHM10}, and the negative-cycle test is its network-flow
form~\cite{Voh11}. The hubs of the upper bounds are points of the tight
span~\cite{Isb64,Dre84}, which enters algorithms through eccentricities and
hyperbolicity~\cite{Lan13,DG21} and enters here as a place to put hubs. The orbit program
of \Cref{sec:midpoint} symmetrizes a covering problem on the Hamming scheme, whose packing
side is Delsarte's linear program and its semidefinite strengthening~\cite{Del73,Sch05};
the hub set $\prod_i([q]\cup\{c_i\})$ is, on two letters, the alphabet $\{0,1,\ast\}$ of
Graham and Pollak's addressing problem~\cite{GP71,Win83}, their third symbol erasing a
coordinate where ours is a midpoint.

%% file: sections/prelim.tex
\section{Preliminaries}\label{sec:prelim}

This section fixes the model and the notation, and isolates the two facts the later
sections call on: how hubs compose across an $\ell_1$ product (\Cref{prop:compose}), and
which sets of pairs a single one can cover (\Cref{prop:cyclic}).

\paragraph{Notation.}
We write $[q]=\{1,\dots,q\}$. All metric spaces are finite and have at least two points,
and we write $X=(V,\delta)$ with $n=|V|$; in a product, $q$ is the number of points of a
factor and $d$ the number of factors. Matrices are indexed by $V\times V$: $J$ is the
all-ones matrix, $I$ the identity, $M\circ N$ the entrywise product,
$\langle M,N\rangle=\sum_{x,y}M_{xy}N_{xy}$, and $\|M\|$ the spectral norm. For
$P\subseteq V\times V$ we write $\mathbf 1_P$ for its indicator matrix.

\subsection{The Model}\label{sec:model}

Cohen, Halperin, Kaplan and Zwick~\cite{CHKZ02} defined a two-hop labeling through the data
stored at each point. We use the equivalent description hub by hub, which turns a labeling
into a covering problem.

\begin{definition}[hubs, labelings, and the three sizes]\label{def:model}
Let $X=(V,\delta)$ be a finite metric space. A \emph{hub} is a quadruple
$h=(A,B,\alpha,\beta)$ with $A,B\subseteq V$, $\alpha\colon A\to\R$ and
$\beta\colon B\to\R$. It is \emph{feasible} if
\[
  \alpha(x)+\beta(y)\ \ge\ \delta(x,y)\qquad\text{for all }(x,y)\in A\times B,
\]
its \emph{tight set} is $T(h)=\{(x,y)\in A\times B:\alpha(x)+\beta(y)=\delta(x,y)\}$, and
its \emph{size} is $|A|+|B|$. A \emph{labeling} of $X$ is a finite family $\mathcal H$ of
feasible hubs whose tight sets cover every ordered pair of distinct points of $V$, and its
size is $\sum_{h\in\mathcal H}(|A_h|+|B_h|)$. We write $\hlst(X)$ for the least size of a
labeling of $X$. Two restrictions of that quantity are the classical models: $\hlts(X)$ is
the least size of a labeling whose hubs draw their two potentials from a single function
$f\colon V\to\R$, one per hub, subject to
\begin{equation}\label{eq:katetov}
  f(x)+f(y)\ \ge\ \delta(x,y)\qquad\text{for all }x,y\in V ,
\end{equation}
and $\hl(X)$ is the least size of a labeling in which each such $f$ is the distance vector
$\delta(\cdot,p)$ of a point $p\in V$. Hubs of the last kind are the ones \cite{CHKZ02}
call proper.
\end{definition}

In the data structure, $x$ stores the value $\alpha_h(x)$ for each hub with $x\in A_h$ and
$\beta_h(x)$ for each hub with $x\in B_h$, and a query on a pair of distinct points $(x,y)$
returns the smallest $\alpha_h(x)+\beta_h(y)$ over the hubs $h$ with $x\in A_h$ and
$y\in B_h$. Feasibility says that this answer is never below $\delta(x,y)$ and the covering
condition says that it is attained, so labelings in the sense above are the exact two-hop
schemes of \cite[Def.~3.3]{CHKZ02}, whose hubs range over an arbitrary finite set and carry
arbitrary real values. A hub carries nothing beyond its two potentials, which is why it has
disappeared from the notation. Size counts stored entries, the unit of
\cite[Rem.~1]{CHKZ02}; the one-sided count of \cite{GRS13} halves it on labelings that
store the same hubs on both sides.

Pairs are ordered, and the diagonal is not required to be covered, though a hub may be
tight there; the proof of \Cref{prop:compose} uses hubs that are. Counting the two sides
separately gives a floor: for $x\in V$ and any $y\ne x$, covering $(x,y)$ places $x$ in
some $A_h$ and covering $(y,x)$ places $x$ in some $B_{h'}$, so
\begin{equation}\label{eq:floor}
  \hlst(X)\ \ge\ 2n .
\end{equation}

The three sizes of \Cref{def:model} satisfy
\begin{equation}\label{eq:chain}
  \hlst(X)\ \le\ \hlts(X)\ \le\ \hl(X),
\end{equation}
each family of labelings sitting inside the previous one, the second inclusion because a
distance vector satisfies \eqref{eq:katetov} by the triangle inequality. For a proper hub
at $p$, feasibility is that triangle inequality and tightness on $(x,y)$ reads
$\delta(x,p)+\delta(p,y)=\delta(x,y)$, that is, $p$ lies between $x$ and $y$ in the sense
of Menger. So $\hl$ is the hub labeling number as it is usually studied~\cite{GRS13}, and
$\hlst$ is what \cite[Def.~3.3]{CHKZ02} allows.

\subsection{The Tight Span}\label{sec:span}

Condition \eqref{eq:katetov} says exactly that the hub with both supports $V$ and both
potentials $f$ is feasible. Its solutions are classical: the pointwise minimal ones form
the tight span $\mathcal T(X)$, Isbell's injective hull of $X$~\cite{Isb64,Dre84}, which is
where the middle size takes its name. Minimality is free. Given a solution $f$ of
\eqref{eq:katetov}, the solutions below $f$ form a compact set, so one of them, say $g$,
minimizes $\sum_xg(x)$; any solution below $g$ is also below $f$, so none lies strictly
below $g$, and $g\in\mathcal T(X)$. Every pair tight for $f$ stays tight for $g$, since
$\delta(x,y)\le g(x)+g(y)\le f(x)+f(y)=\delta(x,y)$ there. Replacing $f$ by $g$ therefore
never increases the size of a labeling, so $\hlts$ is unchanged if its hubs are required to
sit at points of $\mathcal T(X)$.

Points of the span are recognized by partners: if $f$ solves \eqref{eq:katetov} and every
$x$ admits a $y$ with $f(x)+f(y)=\delta(x,y)$, then $f$ is minimal, since lowering $f(x)$
would break \eqref{eq:katetov} at that pair. The distance vector of a point $p\in V$ has
$y=p$ as a partner for every $x$, so proper hubs sit at points of the span. The
constructions of \Cref{sec:hamming,sec:grids} use the criterion in the other direction, to
place hubs at points of the span that are not points of the space.

\subsection{Products, Betweenness, and Constants}\label{sec:products}

The $\ell_1$ product of $X_1=(V_1,\delta_1)$ and $X_2=(V_2,\delta_2)$ is the space
$X_1\oplus X_2$ on $V_1\times V_2$ with
\[
  \delta\bigl((x_1,x_2),(y_1,y_2)\bigr)\ =\ \delta_1(x_1,y_1)+\delta_2(x_2,y_2) .
\]
We write $X^{\oplus d}$ for the $d$-fold power, and $H(d,q)$ for the $d$-fold power of the
uniform metric on $q$ points. This is the Hamming graph $K_q^{\square d}$ with its graph
metric, the hypercube $Q_d$ at $q=2$. A $d$-fold power has $n=q^d$ points, so a size of the
form $r^d$ is the polynomial size $n^{\log_qr}$; the exponential rate in $d$ is what the
rest of the paper computes.

Labelings compose across a product at a constant factor. This is the only general upper
bound the paper needs, and \Cref{sec:hamming,sec:grids} call on it wherever a labeling of a
product is assembled from labelings of its factors.

\begin{proposition}[composition]\label{prop:compose}
For finite metric spaces $X$ and $Y$,
\[
  \hlst(X\oplus Y)\ \le\ 4\,\hlst(X)\,\hlst(Y),
\]
and likewise with $\hlst$ replaced throughout by $\hlts$, or by $\hl$.
\end{proposition}

\begin{proof}
Products of hubs cover the pairs that differ in both coordinates. Each factor's diagonal
has to be bought separately, which at most doubles that factor's labeling, and the two
doublings are the constant.

Let $\mathcal H_1$ and $\mathcal H_2$ be labelings of $X$ and of $Y$ of least size. For
hubs $h=(A_h,B_h,\alpha_h,\beta_h)$ of $X$ and $k=(A_k,B_k,\alpha_k,\beta_k)$ of $Y$, let
$h\otimes k$ be the hub of $X\oplus Y$ with supports $A_h\times A_k$ and $B_h\times B_k$
and potentials $\alpha_h(x_1)+\alpha_k(x_2)$ and $\beta_h(y_1)+\beta_k(y_2)$. Adding the
two feasibility inequalities shows that $h\otimes k$ is feasible, and a sum of two
inequalities is tight exactly when both are, so, identifying
$\bigl((x_1,x_2),(y_1,y_2)\bigr)$ with $\bigl((x_1,y_1),(x_2,y_2)\bigr)$,
\[
  T(h\otimes k)\ =\ T(h)\times T(k) .
\]

Distinct points of $X\oplus Y$ may agree in a coordinate, where the factor labelings offer
nothing. Adjoin therefore to each $\mathcal H_i$ the singleton hubs $(\{z\},\{z\},0,0)$ for
$z\in V_i$, each feasible and tight on the diagonal cell $(z,z)$, and call the enlarged
families $\mathcal H_1^+$ and $\mathcal H_2^+$; their sizes are at most
$\hlst(X)+2|V_1|\le2\hlst(X)$ and $\hlst(Y)+2|V_2|\le2\hlst(Y)$ by \eqref{eq:floor}. Every
ordered pair of distinct points of $X\oplus Y$ is now covered by a product $h\otimes k$
with $h\in\mathcal H_1^+$ and $k\in\mathcal H_2^+$: in each coordinate take a hub of
$\mathcal H_i$ tight on the coordinate pair if the two entries differ, and the singleton
hub at their common value otherwise. Since
$|A_h||A_k|+|B_h||B_k|\le(|A_h|+|B_h|)(|A_k|+|B_k|)$, these products have total size
\[
  \sum_{h\in\mathcal H_1^+}\ \sum_{k\in\mathcal H_2^+}
   \bigl(|A_h||A_k|+|B_h||B_k|\bigr)
  \ \le\ \Bigl(\sum_{h\in\mathcal H_1^+}\bigl(|A_h|+|B_h|\bigr)\Bigr)
         \Bigl(\sum_{k\in\mathcal H_2^+}\bigl(|A_k|+|B_k|\bigr)\Bigr)
  \ \le\ 4\,\hlst(X)\,\hlst(Y).
\]

Both restrictions are inherited. If the potentials of $h$ come from $f$ and those of $k$
from $g$, those of $h\otimes k$ come from $(x_1,x_2)\mapsto f(x_1)+g(x_2)$, which solves
\eqref{eq:katetov} on the product; if moreover $f=\delta_1(\cdot,p_1)$ and
$g=\delta_2(\cdot,p_2)$, that function is $\delta(\cdot,(p_1,p_2))$, and the singleton hub
at $z$ is the proper hub at $z$ with both supports $\{z\}$. Since \eqref{eq:floor} holds a
fortiori for $\hlts$ and $\hl$, the same computation bounds them.
\end{proof}

The construction is standard: it is the mixed product property of Kronecker
products~\cite{AGP23} in the language of labelings, and \cite[\S7]{CHKZ02} apply it to
grids.

By \Cref{def:model} a proper hub can be tight only on pairs whose ends its point lies
between, so the covering problem behind $\hl$ depends on the metric through its betweenness
alone. The class on which we determine the value of Steiner freedom exactly is the one in
which no point lies between two others.

\begin{definition}[betweenness-free]\label{def:bf}
A finite metric space $(V,\delta)$ is \emph{betweenness-free} if there are no three
distinct points $x,y,z$ with $\delta(x,y)+\delta(y,z)=\delta(x,z)$.
\end{definition}

A metric whose largest distance is smaller than twice its smallest positive one is
betweenness-free, since a betweenness equality would make one distance the sum of two
others, hence at least twice the smallest. So is every positively weighted star, the metric
with $\delta(i,j)=\ell_i+\ell_j$ for $i\ne j$ on $[q]$ and all $\ell_i>0$, whose center
lies outside the space: there $\delta(x,y)+\delta(y,z)=\ell_x+2\ell_y+\ell_z$ exceeds
$\delta(x,z)=\ell_x+\ell_z$. The uniform metric on $q$ points is the star with equal arms,
so Hamming graphs and hypercubes are $\ell_1$ products of betweenness-free factors. Three
collinear points are the smallest metric that is not betweenness-free.

Four constants recur, all of them bases of exponentials in the number of factors. For
$q\ge2$,
\begin{equation}\label{eq:constants}
\begin{gathered}
  L_q=\begin{cases}\dfrac52,&q=2\\[6pt]\dfrac{q(2q-3)}{q-1},&q\ge3\end{cases}
  \qquad\qquad
  R_q=\begin{cases}\dfrac32,&q=2\\[6pt]\dfrac{q(q-2)}{q-1},&q\ge3\end{cases}
  \\[10pt]
  \mu_q=\begin{cases}\dfrac{q(q+3)}4,&2\le q\le5\\[6pt]q\sqrt{q-1},&q\ge5\end{cases}
  \qquad\qquad
  \Lz(q)=\frac{q\bigl(2q-1+2\sqrt{q-1}\bigr)}{q+2\sqrt{q-1}} .
\end{gathered}
\end{equation}
$L_q$ and $R_q$ are the two terms of the betweenness-free lower bound (\Cref{thm:Lq}), and
$L_q$ is also the Steiner rate of star products (\Cref{thm:starrate}); $\mu_q$ is the
proper-hub rate of $H(d,q)$ and the ceiling of the invariant of \Cref{sec:invariant}
(\Cref{thm:ceiling}); and $\Lz(q)$ is the floor that holds for every metric on $q$ points
(\Cref{thm:floor}). The two cases of $\mu_q$ agree at $q=5$, and
$\Lz(q)-q=q(q-1)/(q+2\sqrt{q-1})>0$. The constants meet at the small alphabets:
$L_q=\mu_q$ for $q\le3$, while $L_q<\mu_q$ for every $q\ge4$: at $q=4$ that is $20/3<7$,
and for $q\ge5$ it is $L_q/q=2-1/(q-1)<2\le\sqrt{q-1}=\mu_q/q$.

\subsection{Cyclically Monotone Sets}\label{sec:cyclic}

A lower bound has to control what a single hub can do, and the answer is a condition on the
pairs alone.

\begin{definition}\label{def:cyclic}
A set $P\subseteq V\times V$ is \emph{cyclically monotone} if, for every $m\ge1$ and all
pairs $(x_1,y_1),\dots,(x_m,y_m)$ in $P$,
\begin{equation}\label{eq:cyclic}
  \sum_{i=1}^m\delta(x_i,y_i)\ \ge\ \sum_{i=1}^m\delta(x_i,y_{i+1}),
  \qquad y_{m+1}:=y_1 .
\end{equation}
\end{definition}

\begin{proposition}\label{prop:cyclic}
A set $P\subseteq V\times V$ is contained in the tight set of a feasible hub if and only if
it is cyclically monotone. Such a set contains at most one cell of the diagonal.
\end{proposition}

\begin{proof}
Let $P\subseteq T(h)$ for a feasible $h=(A,B,\alpha,\beta)$, and let
$(x_1,y_1),\dots,(x_m,y_m)\in P$. Then $\delta(x_i,y_i)=\alpha(x_i)+\beta(y_i)$ while
$\delta(x_i,y_{i+1})\le\alpha(x_i)+\beta(y_{i+1})$, and the two cyclic sums
$\sum_i\bigl(\alpha(x_i)+\beta(y_i)\bigr)$ and
$\sum_i\bigl(\alpha(x_i)+\beta(y_{i+1})\bigr)$ are equal, which gives \eqref{eq:cyclic}.

Conversely, let $P$ be cyclically monotone and let $A$ and $B$ be its two projections,
which we may assume nonempty. Form the digraph on $B$ carrying, for every ordered pair
$y,y'$ of its points, an arc of weight
\[
  w(y,y')\ =\ \max\bigl\{\delta(x,y')-\delta(x,y)\ :\ (x,y)\in P\bigr\} ,
\]
a maximum over a nonempty set because $y\in B$, and give a directed walk the sum of the
weights of its arcs. A cycle $y_1\to\cdots\to y_m\to y_1$ of positive weight would exhibit
pairs $(x_i,y_i)\in P$ attaining these maxima with
$\sum_i\delta(x_i,y_{i+1})>\sum_i\delta(x_i,y_i)$, against \eqref{eq:cyclic}. No cycle has
positive weight, so a walk never gains by repeating a vertex and
\[
  \beta(y)\ :=\ \max\bigl\{w(W)\ :\ W\ \text{a directed walk ending at }y\bigr\}
\]
is a maximum over finitely many paths, the empty walk giving $\beta\ge0$. Extending an
optimal walk along an arc gives $\beta(y')\ge\beta(y)+w(y,y')$. Now put
$\alpha(x)=\max_{y\in B}\bigl(\delta(x,y)-\beta(y)\bigr)$ for $x\in A$. The hub
$(A,B,\alpha,\beta)$ is feasible by the definition of $\alpha$, and for $(x,y)\in P$ the
arc inequalities read $\delta(x,y)-\beta(y)\ge\delta(x,y')-\beta(y')$ for every $y'\in B$,
so the maximum defining $\alpha(x)$ is attained at $y$ and $(x,y)$ is tight.

Finally, two diagonal cells $(x,x)$ and $(y,y)$ in $P$ give $0\ge2\delta(x,y)$ by
\eqref{eq:cyclic} with $m=2$, hence $x=y$.
\end{proof}

A lower bound must therefore contend with exactly the cyclically monotone sets: the tight
set of every hub is one, and each of them lies in the tight set of some hub. The metric
fixes that family by itself. \Cref{sec:invariant} takes it as the constraint set of a
semidefinite program, and the last sentence of \Cref{prop:cyclic} is what lets the
objective of that program carry mass on the diagonal.

\begin{remark}\label{rem:transport}
\Cref{def:cyclic} says that the pairs of $P$ cannot be rematched cyclically to a larger
total distance. In optimal transport this is $c$-cyclical monotonicity for the cost
$c=-\delta$, the maximizing direction studied by Smith and Knott~\cite{SK92}, and
\Cref{prop:cyclic} is its finite case: the characterization is Rockafellar's~\cite{Roc66},
in Rochet's finite quasi-linear form~\cite{Roc87}. The proof is effective, since $P$ is
cyclically monotone precisely when the digraph it defines has no positive cycle.
\end{remark}

%% file: sections/invariant.tex
\section{The Single-Letter Invariant}\label{sec:invariant}

Every lower bound in this paper comes from one construction. Weight the pairs of a metric
space by a nonnegative kernel and ask how much of that weight a single hub can collect. By
\Cref{prop:cyclic} the answer is a spectral estimate, and it bounds the size of every
labeling. Two facts take it to exponential rates: the estimate multiplies across $\ell_1$
products with no loss (\Cref{thm:tensor}), and a kernel may hold mass on the diagonal until
the product is formed (\Cref{lem:completion}). What remains is a single-letter
optimization, $\Lambda(X)$, whose value bounds the rate of every power of $X$; this section
builds it and determines its range on every finite metric space.

\subsection{The Uniform Metric}\label{sec:uniform}

One example fixes what is at stake. On the uniform metric $U_n$, where all distances are
$1$, hubs off the space save a factor of $n$.

\begin{proposition}\label{prop:uniform}
$\hlst(U_n)=\hlts(U_n)=2n$ and $\hl(U_n)=\Theta(n^2)$.
\end{proposition}

\begin{proof}
The constant function $f\equiv\frac12$ solves \eqref{eq:katetov}, and every point has a
partner, so $f$ is a point of $\mathcal T(U_n)$. The hub with both supports $V$ and both
potentials $f$ is tight on every off-diagonal pair, so $\hlts(U_n)\le2n$, which
\eqref{eq:chain} and \eqref{eq:floor} match from below.

No point of $U_n$ lies between two others, so a proper hub at $p$ is tight only on pairs
having $p$ as an end, and at most $|A_p|+|B_p|$ of those lie in $A_p\times B_p$. Covering
the $n(n-1)$ ordered pairs of distinct points therefore costs at least $n(n-1)$, and the
$n$ proper hubs $(\{p\},V,0,\delta(p,\cdot))$ cover all of them at total cost $n(n+1)$.
\end{proof}

The proper side is folklore. The hub that delivers the other two lies at distance $\frac12$
from every point and has no counterpart in the space, which is the phenomenon the rest of
the paper prices.

\subsection{Kernels and the Packing Bound}\label{sec:kernels}

A lower bound has to hold against every labeling at once. Weighting the pairs reduces that
to a single hub, whose tight set can collect only so much weight for its size. From here on
$X=(V,\delta)$ is a metric space on $q$ points.

\begin{definition}\label{def:kernel}
A \emph{kernel} on $X$ is a symmetric matrix $K\ge0$ indexed by $V$, with the diagonal
allowed. Its \emph{mask norm} is
\begin{equation}\label{eq:CX}
  C_X(K)\ =\ \max\bigl\{\,\|K\circ\mathbf 1_P\|\ :\ P\subseteq V\times V
  \ \text{cyclically monotone}\,\bigr\}.
\end{equation}
\end{definition}

Masking a nonnegative matrix cannot raise its norm, since $0\le M\le M'$ entrywise gives
$|x|^{\top}M|y|\le|x|^{\top}M'|y|$ for all $x,y$. So the maximum in \eqref{eq:CX} is
attained on maximal cyclically monotone sets, which by \Cref{prop:cyclic} are the tight
sets of feasible hubs.

\begin{lemma}\label{lem:packing}
Let $K$ be a kernel on $X$ with zero diagonal. Every labeling of $X$ has size at least
$2\langle J,K\rangle/C_X(K)$.
\end{lemma}

\begin{proof}
Let $\mathcal H$ be a labeling. Every ordered pair of distinct points lies in some tight
set and $K$ vanishes on the diagonal, so
\[
  \langle J,K\rangle\ \le\ \sum_{h\in\mathcal H}\ \sum_{(x,y)\in T(h)}K_{xy}\,.
\]
Fix $h$. Its inner sum is $\mathbf 1_{A_h}^{\top}(K\circ\mathbf 1_{T(h)})\mathbf 1_{B_h}$,
at most $\|K\circ\mathbf 1_{T(h)}\|\sqrt{|A_h||B_h|}$, and $T(h)$ is cyclically monotone,
so the sum is at most $C_X(K)\sqrt{|A_h||B_h|}\le C_X(K)(|A_h|+|B_h|)/2$. Summing over
$\mathcal H$ gives the claim.
\end{proof}

Every bound below is this packing bound applied to one kernel. The work is in producing
kernels whose mask norm stays under control on a whole family of spaces at once, and in
making the control survive products.

\subsection{Tensorization}\label{sec:tensor}

A hub of a product need not decompose, and its tight set need not be a product of factor
tight sets: on $Q_2$ the hub with supports $\{00,11\}$ and $\{01,10\}$ and potentials $0$
and $1$ is tight on all four of its pairs, and either coordinate projection of that tight
set is the full square, which is not cyclically monotone. What does separate is the
tightness condition itself, along a fibration that crosses the two coordinates.

\begin{theorem}\label{thm:tensor}
Let $K_1$ be a kernel on $X_1$ and $K_2$ a kernel on $X_2$. Then
\[
  C_{X_1\oplus X_2}(K_1\otimes K_2)\ \le\ C_{X_1}(K_1)\,C_{X_2}(K_2).
\]
\end{theorem}

\begin{proof}
Pair the row's second coordinate with the column's first. Along each fiber so formed the
tightness condition has an extreme value; the extremizers are tight sets of genuine factor
hubs, and the two families assemble into a product of block-diagonal matrices.

By \Cref{prop:cyclic} and the monotonicity of the norm it is enough to bound
$\|(K_1\otimes K_2)\circ\mathbf 1_{T(h)}\|$ for a feasible hub $h=(\mathcal A,\mathcal
B,\alpha,\beta)$ of $X_1\oplus X_2$. Write points as $(x,u)$ with $x\in V_1$ and $u\in V_2$,
so that feasibility reads
\begin{equation}\label{eq:prodfeas}
  \alpha(x,u)+\beta(y,v)\ \ge\ \delta_1(x,y)+\delta_2(u,v),
  \qquad (x,u)\in\mathcal A,\ (y,v)\in\mathcal B .
\end{equation}
Put $A_u=\{x:(x,u)\in\mathcal A\}$ and $B_y=\{v:(y,v)\in\mathcal B\}$, let $\mathcal I$ be
the set of pairs $(u,y)$ with both nonempty, and for $(u,y)\in\mathcal I$ set
\[
  F_{uy}(x)=\alpha(x,u)-\delta_1(x,y),\qquad G_{uy}(v)=\delta_2(u,v)-\beta(y,v),
\]
with $\underline F_{uy}=\min_{A_u}F_{uy}$ and $\overline G_{uy}=\max_{B_y}G_{uy}$.
Rearranged, \eqref{eq:prodfeas} says exactly that $F_{uy}(x)\ge G_{uy}(v)$, so
\[
  F_{uy}(x)\ \ge\ \underline F_{uy}\ \ge\ \overline G_{uy}\ \ge\ G_{uy}(v),
\]
and a pair is tight precisely when this chain collapses:
\begin{equation}\label{eq:tightsplit}
  \bigl((x,u),(y,v)\bigr)\in T(h)
  \iff F_{uy}(x)=\underline F_{uy}\ \text{ and }\ \underline F_{uy}=\overline G_{uy}
  \ \text{ and }\ G_{uy}(v)=\overline G_{uy} .
\end{equation}

The two outer conditions are factor tightness. For fixed $u$, the hub of $X_1$ with
supports $A_u$ and $\{y:B_y\ne\varnothing\}$ and potentials $x\mapsto\alpha(x,u)$ and
$y\mapsto-\underline F_{uy}$ has slack $F_{uy}(x)-\underline F_{uy}\ge0$, so it is feasible
and its tight set is $\{(x,y):F_{uy}(x)=\underline F_{uy}\}$; let $R^u$ be $K_1$ masked by
that set, so $\|R^u\|\le C_{X_1}(K_1)$. Symmetrically, for fixed $y$ the hub of $X_2$ with
supports $\{u:A_u\ne\varnothing\}$ and $B_y$ and potentials $u\mapsto\overline G_{uy}$ and
$v\mapsto\beta(y,v)$ has slack $\overline G_{uy}-G_{uy}(v)\ge0$, and $S^y$, the mask of
$K_2$ by its tight set, has $\|S^y\|\le C_{X_2}(K_2)$.

Now assemble $R\in\R^{\mathcal A\times\mathcal I}$, a diagonal $D$ on $\mathcal I$, and
$S\in\R^{\mathcal I\times\mathcal B}$ by
\[
  R_{(x,u),(u',y)}=\mathbf 1\{u=u'\}\,R^u_{xy},\qquad
  D_{(u,y)}=\mathbf 1\{\underline F_{uy}=\overline G_{uy}\},\qquad
  S_{(u,y),(y',v)}=\mathbf 1\{y=y'\}\,S^y_{uv}.
\]
For fixed $(x,u)$ and $(y,v)$ the only index of $\mathcal I$ contributing to $RDS$ is
$(u,y)$, so \eqref{eq:tightsplit} is the entrywise identity
$RDS=(K_1\otimes K_2)\circ\mathbf 1_{T(h)}$. Distinct $u$ occupy disjoint rows and disjoint
columns of $R$, and distinct $y$ likewise in $S$, so $R=\bigoplus_uR^u$ and
$S=\bigoplus_yS^y$ are direct sums. Hence $\|R\|\le C_{X_1}(K_1)$, $\|D\|\le1$ and
$\|S\|\le C_{X_2}(K_2)$, and the identity finishes the proof.
\end{proof}

\begin{remark}\label{rem:gamma2}
Constraining a matrix through a family of masked spectral norms is the filtered $\gamma_2$
norm of Lee, Mittal, Reichardt, Špalek and Szegedy~\cite[Def.~3.1]{LMRSS11}, ordinary
$\gamma_2$ being the case of a single mask, and exact multiplicativity of such norms under
$\otimes$ is known~\cite[Thm.~17]{LSS08}. Neither applies here: those filter families are
fixed under tensoring, while the cyclically monotone sets of $X_1\oplus X_2$ are not the
products of those of $X_1$ and $X_2$, as the hub above already shows. The crossed fibration
is what replaces the missing product structure.
\end{remark}

\subsection{Diagonal Completion and the Invariant}\label{sec:completion}

\Cref{lem:packing} asks for a zero diagonal, and where that constraint is imposed decides
what a kernel can see. Imposed before a product is formed, it strips the diagonal from every
letter; imposed afterwards, it costs one deletion, and the diagonal of a $d$-fold tensor is
exponentially thinner than the tensor. Nothing in \Cref{thm:tensor} reads the diagonal, so
the second order is available.

\begin{lemma}\label{lem:completion}
For $i=1,\dots,d$ let $K_i$ be a kernel on $X_i$ with $C_{X_i}(K_i)\le1$. Then
\[
  \hlst(X_1\oplus\cdots\oplus X_d)\ \ge\
  2\Bigl(\prod_{i=1}^d\langle J,K_i\rangle-\prod_{i=1}^d\operatorname{tr}K_i\Bigr).
\]
\end{lemma}

\begin{proof}
By \Cref{thm:tensor} and induction on $d$, the tensor $K_1\otimes\cdots\otimes K_d$ has
mask norm at most $1$ on the product. Let $W$ agree with it off the diagonal and vanish on
the diagonal. Then $W$ is a kernel with zero diagonal, and
$0\le W\circ\mathbf 1_P\le(\bigotimes_iK_i)\circ\mathbf 1_P$ entrywise for every $P$, so
$W$ has mask norm at most $1$ as well. Both $\langle J,\cdot\rangle$ and the trace are
multiplicative under $\otimes$, so $\langle J,W\rangle=\prod_i\langle
J,K_i\rangle-\prod_i\operatorname{tr}K_i$, and \Cref{lem:packing} applies.
\end{proof}

On two points the deletion is the entire bound. Take $X$ with two points at distance $1$
and
\[
  K\ =\ \tfrac34I+\tfrac12(J-I).
\]
A mask carries at most one diagonal cell by \Cref{prop:cyclic}, so it lies either inside
the off-diagonal, where $K$ masks to a matrix of norm $\frac12$, or inside the cross at a
point, where $K$ masks to $\bigl(\begin{smallmatrix}3/4&1/2\\1/2&0\end{smallmatrix}\bigr)$,
whose eigenvalues are $1$ and $-\frac14$. So $C_X(K)=1$, and since $\langle
J,K\rangle=\frac52$ and $\operatorname{tr}K=\frac32$, \Cref{lem:completion} gives
\[
  \hlst(Q_d)\ \ge\ 2\bigl[(\tfrac52)^d-(\tfrac32)^d\bigr].
\]
A zero diagonal imposed first leaves $\kappa(J-I)$, whose masks all have norm $\kappa$, so
its value is at most $2$ and its bound is $2\cdot2^d$, which is \eqref{eq:floor} and nothing
more.

\begin{definition}\label{def:lambda}
\[
  \Lambda(X)\ =\ \max\bigl\{\,\langle J,K\rangle\ :\ K\ \text{a kernel on }X
  \ \text{with}\ C_X(K)\le1\,\bigr\}.
\]
\end{definition}

The maximum is attained: a single cell is cyclically monotone, so feasibility forces
$K_{xy}\le1$ for every pair and the feasible set is compact.

\begin{proposition}\label{prop:rate}
For every finite metric space $X$,
\[
  \liminf_{d\to\infty}\hlst(X^{\oplus d})^{1/d}\ \ge\ \Lambda(X).
\]
\end{proposition}

\begin{proof}
Let $K$ attain $\Lambda(X)$ and put $\Delta=\operatorname{tr}K$. If $K$ has a nonzero
off-diagonal entry then $\Delta<\Lambda(X)$, and \Cref{lem:completion} gives
$\hlst(X^{\oplus d})\ge2\Lambda(X)^d\bigl(1-(\Delta/\Lambda(X))^d\bigr)$, whose $d$-th root
tends to $\Lambda(X)$. Otherwise $K$ is diagonal, so $\Lambda(X)=\Delta\le q$;
since $(J-I)/(q-1)$ has mask norm at most $1$ and value $q$, this forces $\Lambda(X)=q$, and
the argument applies to that kernel, whose trace is $0$.
\end{proof}

So $\Lambda$ bounds the rate of every power of $X$. The rest of the section bounds
$\Lambda$ itself, from both sides, with no hypothesis on $X$.

\subsection{The Floor}\label{sec:floor}

A kernel must survive two mask geometries: those lying inside the off-diagonal, and those
carrying a diagonal cell, of which \Cref{prop:cyclic} allows at most one. One kernel
balances the two on every metric at once, and the balance point is what fixes $\Lz(q)$.

\begin{theorem}\label{thm:floor}
$\Lambda(X)\ge\Lz(q)$ for every metric space $X$ on $q\ge2$ points.
\end{theorem}

\begin{proof}
Write $s=\sqrt{q-1}$ and take
\[
  K\ =\ c\,(J-I)+t\,I,\qquad c=\frac1{s(s+1)},\qquad t=\frac{s^2+s+1}{(s+1)^2}.
\]
A mask with no diagonal cell gives $K\circ\mathbf 1_P\le c(J-I)$ entrywise, of norm
$c(q-1)=s/(s+1)<1$. A mask carrying the diagonal cell at $p$ gives $K\circ\mathbf 1_P\le
c(J-I)+tE_{pp}$ entrywise, and the vector $w$ with $w_p=s+1$ and $w_x=1$ elsewhere is fixed
by that matrix: its $p$-th coordinate is $t(s+1)+cs^2=(s^2+s+1+s)/(s+1)=s+1$, and its other
coordinates are $c\bigl((s+1)+(q-2)\bigr)=c\,s(s+1)=1$. Since $w$ is positive and the
matrix is symmetric and nonnegative, Perron--Frobenius makes its norm exactly $1$. By
\Cref{prop:cyclic} these two cases exhaust the masks, so $C_X(K)\le1$. Finally
\[
  \langle J,K\rangle\ =\ q\bigl(c(q-1)+t\bigr)
  \ =\ q\Bigl(\frac{s}{s+1}+\frac{s^2+s+1}{(s+1)^2}\Bigr)
  \ =\ \frac{q(2s^2+2s+1)}{(s+1)^2}\ =\ \Lz(q). \qedhere
\]
\end{proof}

With \Cref{prop:rate}, no $\ell_1$ power is free: on $n=q^d$ points, every $d$-fold power
of every $q$-point metric has $\hlst\ge n^{1+c_q-o(1)}$ with $c_q=\log_q(\Lz(q)/q)>0$,
since $\Lz(q)-q=q(q-1)/(q+2s)>0$. Comparing with \eqref{eq:constants}, $\Lz(2)=L_2=\frac52$,
while $\Lz(q)<L_q$ for $q\ge3$, that comparison reducing to $2(q-2)\sqrt{q-1}>1$.

\subsection{The Ceiling}\label{sec:ceiling}

The proper hub that defines the third size in \Cref{def:model} also constrains every
kernel, and the constraint is sharp enough to cap $\Lambda$.

\begin{theorem}\label{thm:ceiling}
Let $X$ be a metric space on $q\ge2$ points. Then $\Lambda(X)\le\mu_q$, with equality when
$q=2$ and with strict inequality when $q\ge4$.
\end{theorem}

\begin{proof}
The cross at a point caps each row of $K$; the caps sum to $\mu_q$; equality forces a single
kernel, and one further mask refutes it.

Let $K$ be a kernel with $C_X(K)\le1$ and fix $p\in V$. The proper hub at $p$ is tight on
the cross at $p$, so the cross is a mask. Masking $K$ by it leaves $K_{pp}$ at $(p,p)$ and
the vector $v_p=(K_{py})_{y\ne p}$ along the rest of row and column $p$, a matrix of norm
$\frac12\bigl(K_{pp}+\sqrt{K_{pp}^2+4\|v_p\|^2}\bigr)$, so feasibility forces
$\|v_p\|^2\le1-K_{pp}$. With Cauchy--Schwarz on each row,
\begin{equation}\label{eq:ceilchain}
\begin{split}
  \langle J,K\rangle=\sum_{p}\Bigl(K_{pp}+\sum_{y\ne p}K_{py}\Bigr)
  &\ \le\ \sum_p\Bigl(K_{pp}+\sqrt{(q-1)(1-K_{pp})}\Bigr)\\
  &\ \le\ q\max_{0\le\tau\le1}\Bigl(\tau+\sqrt{(q-1)(1-\tau)}\Bigr)\ =\ \mu_q,
\end{split}
\end{equation}
the maximum being attained at $\tau=(5-q)/4$ for $q\le5$ and at $\tau=0$ for $q\ge5$.

Equality throughout \eqref{eq:ceilchain} forces the kernel. The last step requires
$K_{pp}=\tau$ at every $p$, and Cauchy--Schwarz requires the off-diagonal entries of each
row to agree, so $K=\tau I+\kappa(J-I)$ with
$\kappa=\sqrt{(1-\tau)/(q-1)}=1/\max\{2,\sqrt{q-1}\}$.

At $q=2$ that kernel is $\frac34I+\frac12(J-I)$, shown feasible in \Cref{sec:completion},
so $\Lambda=\mu_2=\frac52$ on every two-point space.

For $q\ge4$ one mask refutes it, and every metric supplies that mask. Let $\{j,k\}$
minimize the Gromov product $(x\cdot y)_p=\frac12\bigl(\delta(x,p)+\delta(p,y)-\delta(x,y)
\bigr)$ over pairs of distinct points other than $p$, and let $\gamma$ be its minimum value.
The hub with both supports $V$ and both potentials equal to $\gamma$ at $p$ and to
$\delta(\cdot,p)-\gamma$ elsewhere is feasible: pairs meeting $p$ have slack $0$, the cell
$(p,p)$ has slack $2\gamma\ge0$, and a pair $x,y\ne p$ has slack
$2\bigl((x\cdot y)_p-\gamma\bigr)\ge0$, the case $x=y$ included because
$\gamma\le(x\cdot z)_p\le\delta(x,p)$ for any third point $z$. Its tight set contains the
cross at $p$ together with $(j,k)$ and $(k,j)$, so the star at $p$ with the edge $jk$
adjoined is a mask. That graph has a Perron vector constant on $\{j,k\}$ and constant off
$\{p,j,k\}$, say with values $a$, $b$ and $c$ there, so its spectral radius satisfies
$\sigma a=2b+(q-3)c$, $\sigma b=a+b$ and $\sigma c=a$. Eliminating $a$ and $b$ makes
$\sigma_q$ the largest root of
\[
  x^3-x^2-(q-1)x+(q-3),
\]
a polynomial taking the value $3-q$ at $2$ and the value $-2$ at $\sqrt{q-1}$. For $q\ge4$
both values are negative, so $\sigma_q>\max\{2,\sqrt{q-1}\}=1/\kappa$. The mask is
diagonal-free, so $\|K\circ\mathbf 1_P\|=\kappa\sigma_q>1$ and $K$ is infeasible.
\end{proof}

The refuting mask is balanced exactly at $q=3$, where the polynomial vanishes at $2$;
\Cref{sec:hamming} shows that the ceiling is attained there, and \Cref{sec:grids}
that it is attained by nothing else on three points. That ceiling is the rate proper hubs
achieve on the Hamming graph $H(d,q)$ (\Cref{thm:starrate}), reached here by a computation
that shares nothing with that one, so the method stops exactly where proper hubs stand: on
the betweenness-free cone the invariant reaches it for $q\le3$ and falls short from $q=4$,
and the shortfall is the gain.

%% file: sections/hamming.tex
\section{Betweenness-Free Products}\label{sec:hamming}

On the cone of $\ell_1$ products of betweenness-free factors both ends of \eqref{eq:chain}
can be computed. The invariant of \Cref{sec:invariant} takes the value $L_q$ there and no
less, and the positively weighted stars are the only bases where it is that small. Proper
hubs see such a product through its alphabet alone, so their rate is the Hamming one,
$\mu_q$, and on star products a dictionary of midpoints matches the lower bound. Since
$L_q=\mu_q$ for $q\le3$ and $L_q<\mu_q$ for $q\ge4$, the value of Steiner freedom on this
cone is decided by the alphabet alone.

\subsection{The Floor on the Cone}\label{sec:cone}

In \Cref{sec:floor} a mask carrying a diagonal cell was free to spread over the rest of the
space. Betweenness-freeness collapses it onto a cross, and the kernel balancing the two
geometries then carries more value.

\begin{lemma}\label{lem:collapse}
Let $X$ be betweenness-free and let $P$ be a cyclically monotone set containing the
diagonal cell $(p,p)$. Then $P$ lies in the cross at $p$.
\end{lemma}

\begin{proof}
By \Cref{prop:cyclic}, $P\subseteq T(h)$ for a feasible hub $h=(A,B,\alpha,\beta)$ with
$p\in A\cap B$ and $\alpha(p)+\beta(p)=0$. Let $x\in A$ and $y\in B$ with $x,y\ne p$.
Feasibility at $(x,p)$ and at $(p,y)$ gives $\alpha(x)\ge\delta(x,p)-\beta(p)$ and
$\beta(y)\ge\delta(p,y)-\alpha(p)$, so
\[
  \alpha(x)+\beta(y)\ \ge\ \delta(x,p)+\delta(p,y)-\bigl(\alpha(p)+\beta(p)\bigr)
  \ =\ \delta(x,p)+\delta(p,y).
\]
For $x\ne y$ the right side exceeds $\delta(x,y)$ because $p$ lies between no two points,
and for $x=y$ it equals $2\delta(x,p)>0=\delta(x,x)$. Either way $(x,y)\notin T(h)$.
\end{proof}

\begin{theorem}\label{thm:Lq}
Every betweenness-free metric space on $q\ge2$ points has $\Lambda(X)\ge L_q$, and for
arbitrary betweenness-free $X_1,\dots,X_d$ on $q$ points each,
\[
  \hlst(X_1\oplus\cdots\oplus X_d)\ \ge\ 2\bigl(L_q^{\,d}-R_q^{\,d}\bigr).
\]
\end{theorem}

\begin{proof}
For $q\ge3$ take the kernel
\[
  K\ =\ \frac{(q-2)I+(J-I)}{q-1},
\]
and for $q=2$ the kernel $\frac34I+\frac12(J-I)$ of \Cref{sec:completion}, whose value and
trace are $L_2$ and $R_2$. A mask missing the diagonal is dominated entrywise by
$(J-I)/(q-1)$, of norm $1$. A mask meeting the diagonal at $p$ lies in the cross at $p$ by
\Cref{lem:collapse}, so it is dominated by the arrowhead with $(q-2)/(q-1)$ at $(p,p)$ and
$1/(q-1)$ along the rest of row and column $p$. Its arm vector has
$\|v_p\|^2=(q-1)^{-1}=1-K_{pp}$, which by the arrowhead computation of \Cref{thm:ceiling}
is exactly the condition for norm $1$. Hence $C_X(K)\le1$, and
\[
  \langle J,K\rangle=\frac{q(q-2)+q(q-1)}{q-1}=L_q,\qquad
  \operatorname{tr}K=\frac{q(q-2)}{q-1}=R_q .
\]
Both values read only $q$, so \Cref{lem:completion} applies to any product of
betweenness-free factors on $q$ points, one such kernel per factor.
\end{proof}

The coefficient is forced rather than chosen. For $K_u=uI+(J-I)$ the same two geometries
give $C_X(K_u)\le\max\bigl\{q-1,\tfrac12(u+\sqrt{u^2+4(q-1)})\bigr\}$, and the guaranteed
value $q(u+q-1)$ divided by that bound is largest where the two branches meet: the
diagonal-free masks and the crosses cost the same there, and any other coefficient wastes
one of them. For $q\ge3$ that balance point is $u=q-2$. At $q=2$ the first branch is the
smaller one and the balance moves to $u=3/2$, where $C_X(K_{3/2})=2$ and the normalized
kernel $\frac12K_{3/2}$ is the one \Cref{sec:completion} used on the cube.

Stars are the bases where the first geometry is as large as it can be, and that is what
makes them extremal.

\begin{theorem}\label{thm:unique}
Let $q\ge4$ and let $X$ be betweenness-free on $q$ points. If $X$ is not a positively
weighted star, then $\Lambda(X)>L_q$.
\end{theorem}

\begin{proof}
Stars are exactly the bases whose whole off-diagonal is one mask; a non-star loses spectral
radius there, and the loss can be spent on the diagonal.

If $J-I$ is cyclically monotone there are potentials with $\alpha_i+\beta_j=\delta(i,j)$
for all $i\ne j$, so reading $\alpha_i+\beta_j+\alpha_k+\beta_l$ two ways gives
$\delta(i,j)+\delta(k,l)=\delta(i,l)+\delta(k,j)$ for distinct $i,j,k,l$. Then
$\ell_i=\frac12\bigl(\delta(i,j)+\delta(i,k)-\delta(j,k)\bigr)$ does not depend on
$\{j,k\}$, and $\delta(i,j)=\ell_i+\ell_j$ with every $\ell_i>0$ by betweenness-freeness.
Conversely the center hub $\alpha=\beta=\ell$ of a positively weighted star is feasible and
tight on all of $J-I$.

So for a non-star every diagonal-free mask is a proper subset of the off-diagonal. Fix one,
$P$. Then $\mathbf 1_P^{\top}\mathbf 1_P\le(J-I)^2=(q-2)J+I$ entrywise, the right side is
positive and therefore irreducible, and the two differ in trace, $|P|<q(q-1)$. Strict
Perron--Frobenius monotonicity gives $\|\mathbf 1_P\|<q-1$, and as there are finitely many
masks, $\rho:=\max_P\|\mathbf 1_P\|<q-1$.

That slack buys value. Put $u=\min\{\tfrac12,(q-1)^{-1/2},\rho^{-1}\}$ and
\[
  K\ =\ u\,(J-I)+\bigl(1-u^2(q-1)\bigr)I ,
\]
whose diagonal is nonnegative. A diagonal-free mask gives $\|K\circ\mathbf 1_P\|=u\|\mathbf
1_P\|\le u\rho\le1$; a mask meeting the diagonal lies in a cross by \Cref{lem:collapse}, and
there the arrowhead condition of \Cref{thm:ceiling} reads $u^2(q-1)\le1-K_{pp}$, an equality
by construction. So $C_X(K)\le1$ and $\Lambda(X)\ge\langle J,K\rangle=q\,g(u)$ with
$g(z)=1+(q-1)z-(q-1)z^2$, which increases on $[0,\frac12]$. Since $q\ge4$ makes each of
$\frac12$, $(q-1)^{-1/2}$ and $\rho^{-1}$ exceed $(q-1)^{-1}$, we get $u>(q-1)^{-1}$ and
\[
  \Lambda(X)\ \ge\ q\,g(u)\ >\ q\,g\Bigl(\frac1{q-1}\Bigr)
  =q\Bigl(2-\frac1{q-1}\Bigr)=L_q. \qedhere
\]
\end{proof}

Every betweenness-free metric on at most three points is a positively weighted star: on two
points take any $\ell_1,\ell_2>0$ summing to the distance, and on three the lengths
$\ell_i=\frac12\bigl(\delta(i,j)+\delta(i,k)-\delta(j,k)\bigr)$ reconstruct $\delta$ and are
positive exactly by betweenness-freeness. So at $q=3$ there is nothing to separate, and
\Cref{thm:Lq} gives $\Lambda\ge L_3=9/2=\mu_3$, which with \Cref{thm:ceiling} forces
equality. That settles the case the ceiling theorem left open: it is attained at $q=3$ and
at no larger alphabet.

\subsection{Proper Hubs See Only the Alphabet}\label{sec:oblivious}

\begin{proposition}\label{prop:oblivious}
Let $D=\delta_1\oplus\cdots\oplus\delta_d$ with every $\delta_i$ betweenness-free on $[q]$.
For points $x,y,h$ of $D$,
\[
  D(x,h)+D(h,y)=D(x,y)\iff h_i\in\{x_i,y_i\}\ \text{for every }i .
\]
The proper-hub covering problem of $D$ is therefore the one of $H(d,q)$, at every finite
$d$ and for every choice of factors and lengths.
\end{proposition}

\begin{proof}
The slack is $\sum_i[\delta_i(x_i,h_i)+\delta_i(h_i,y_i)-\delta_i(x_i,y_i)]$, a sum of
nonnegative terms, so it vanishes exactly when each term does. At a coordinate with
$x_i=y_i$ the term is $2\delta_i(x_i,h_i)$, zero only for $h_i=x_i$; at a coordinate with
$x_i\ne y_i$ it vanishes exactly when $h_i$ lies between $x_i$ and $y_i$, which for a
betweenness-free factor means $h_i\in\{x_i,y_i\}$. The resulting relation reads the
coordinate patterns and nothing else.
\end{proof}

So $\hl$ is constant along the whole cone at each $d$, and computing it once on $H(d,q)$
computes it everywhere. The lengths survive only in the Steiner sizes, and
\Cref{sec:midpoint} removes them there too.

\subsection{Midpoints and the Orbit Program}\label{sec:midpoint}

Each star factor contributes one point of the tight span that the space itself does not
contain: its center. Adjoining them gives a dictionary of $(q+1)^d$ hubs whose covering
problem can be solved exactly in the exponent.

\begin{lemma}\label{lem:span}
Let $D=\delta_1\oplus\cdots\oplus\delta_d$ with $\delta_i$ a positively weighted star on
$[q]$ with center $c_i$, and let $\mathcal M=\prod_i\bigl([q]\cup\{c_i\}\bigr)$. Every
$h\in\mathcal M$ is a point of $\mathcal T(D)$, and $h$ is tight on $(x,y)$ exactly when,
in every coordinate, either $h_i\in\{x_i,y_i\}$ or $h_i=c_i$ and $x_i\ne y_i$.
\end{lemma}

\begin{proof}
In one star with center $c$, a leaf $z$ is tight on a pair $(w,w')$ exactly when
$z\in\{w,w'\}$, the two distances summing to $\delta(w,w')+2\ell_z$ otherwise; and $c$ is
tight on $(w,w')$ exactly when $w\ne w'$, since $\ell_w+\ell_{w'}=\delta(w,w')$ while
$2\ell_w>0$. Neither condition reads a length. Summing the coordinate slacks as in
\Cref{prop:oblivious} gives the tightness rule.

For the span, $f=D(\cdot,h)$ satisfies \eqref{eq:katetov} by the triangle inequality, and a
partner for $x$ is obtained coordinatewise: take $y_i=h_i$ where $h_i$ is a leaf, and any
$y_i\ne x_i$ where $h_i=c_i$. Each coordinate then contributes
$\delta_i(x_i,h_i)+\delta_i(h_i,y_i)=\delta_i(x_i,y_i)$, so $f(x)+f(y)=D(x,y)$ and $f$ is
minimal by the criterion of \Cref{sec:span}.
\end{proof}

By \Cref{lem:span,prop:oblivious} the covering problem over $\mathcal M$ depends only on
$q$ and $d$, so we may compute it on $H(d,q)$. Write
$\Gamma_h$ for the graph on $[q]^d$ joining $x$ to $y$ when $h$ is tight on the pair, let
$E_k(\Gamma_h)$ be its edges between points at Hamming distance $k$, and put
\[
  N_k=\tfrac12q^d\binom dk(q-1)^k,\qquad
  D_k=\max_{h\in\mathcal M,\ S\subseteq[q]^d}\frac{|E_k(\Gamma_h[S])|}{|S|} .
\]
$N_k$ counts the pairs a labeling must cover at distance $k$ and $D_k$ is the most any
single hub can cover per unit of size, so the ratio $N_k/D_k$ is the obstruction at that
distance. Both have closed forms. Call a coordinate of $h$ a center coordinate when
$h_i=c_i$ and a leaf coordinate otherwise, and write $\sigma(x)$ for the number of leaf
coordinates at which $x$ differs from $h$.

\begin{lemma}\label{lem:layer}
Let $h$ have $r$ center and $m=d-r$ leaf coordinates, and put $\ell=k-r$. The layer
$E_k(\Gamma_h)$ is empty unless $0\le\ell\le m$, and otherwise it is the vertex-disjoint
union of the pieces $\Pi_i$, $0\le i\le\lfloor\ell/2\rfloor$, spanned by the points with
$\sigma=i$ and $\sigma=\ell-i$, the piece $\Pi_i$ being biregular with degrees $d_i$ and
$d_{\ell-i}$ for $d_i=\binom{m-i}{\ell-i}(q-1)^{r+\ell-i}$. The densest induced subgraph of
$\Pi_i$ has density $\bigl(d_i^{-1}+d_{\ell-i}^{-1}\bigr)^{-1}$, largest at
$i=\lfloor\ell/2\rfloor$, and for $q\ge3$
\[
  D_k\ =\ \max_{a+b=k,\ a\le b}\ \tfrac12\binom{d-b}{a}(q-1)^{b} .
\]
\end{lemma}

\begin{proof}
By \Cref{lem:span} a center coordinate forces $x_i\ne y_i$, so all $r$ of them contribute to
the Hamming distance; a leaf coordinate contributes $0$ when $x_i=y_i=h_i$ and $1$ when
exactly one of $x_i,y_i$ equals $h_i$. So an edge of $E_k(\Gamma_h)$ has
$\sigma(x)+\sigma(y)=\ell$ with disjoint leaf deviation sets, which gives the decomposition
and its disjointness. For the degree, fix $x$ with $\sigma(x)=i$: a neighbor is obtained by
choosing the $\ell-i$ leaf coordinates where $y$ deviates, from the $m-i$ where $x$ does
not, a deviating value at each, and a value differing from $x_i$ at each center coordinate,
which is $d_i$ choices and does not depend on $x$.

In a bipartite graph with left degree $d_L$ and right degree $d_R$, an induced subgraph on
$s_L+s_R$ vertices with $e$ edges has $e\le d_Ls_L$ and $e\le d_Rs_R$, hence density at most
$(d_L^{-1}+d_R^{-1})^{-1}$, attained by the whole piece; a regular piece obeys the same
bound with $d_L=d_R$. Since the pieces are vertex-disjoint, the densest subgraph of the
layer is the densest piece. To locate it, put
$u_i=\bigl[\binom{m-\ell+i}{i}(q-1)^i\bigr]^{-1}$, so
that $u_i=(q-1)^r/d_{\ell-i}$ and the density of $\Pi_i$ is $(q-1)^r/(u_i+u_{\ell-i})$. From
$\binom{m-\ell+i+1}{i+1}=\binom{m-\ell+i}{i}\frac{m-\ell+i+1}{i+1}$,
\[
  \frac{u_{i+1}}{u_i}=\frac1{q-1}\cdot\frac{i+1}{m-\ell+i+1}\ \le\
  \frac{\ell+1}{(q-1)(m+1)}\ \le\ 1 ,
\]
and the ratio is nondecreasing in $i$. So $u$ is nonincreasing and its forward differences
$u_i-u_{i+1}=u_i(1-u_{i+1}/u_i)$, being products of two nonnegative nonincreasing sequences,
are nonincreasing as well. For $i<\ell/2$ one has $i\le\ell-i-1$, so
$(u_i+u_{\ell-i})-(u_{i+1}+u_{\ell-i-1})=(u_i-u_{i+1})-(u_{\ell-i-1}-u_{\ell-i})\ge0$: the
sum falls towards the center and the density rises.

For even $\ell=2a$ both central degrees equal $\binom{m-a}{a}(q-1)^{r+a}$, which with
$b=r+a$ and $k=a+b$ is $\binom{d-b}{a}(q-1)^b$, and the density is half of it. For odd
$\ell=2a+1$ the central pieces have degrees $d_{a+1}=\binom{d-b}{a}(q-1)^{b-1}$ and
$d_a=\binom{d-b+1}{a+1}(q-1)^{b}$ with $b=r+a+1$, so with $R=d_a/d_{a+1}=(q-1)\frac{d-b+1}
{a+1}$ the density is $d_{a+1}R/(R+1)$. The hub with one further center coordinate at the
same $k$ has the same $b$ and even $\ell$, hence density $(q-1)d_{a+1}/2$, and
\[
  \frac{d_{a+1}R/(R+1)}{(q-1)d_{a+1}/2}=\frac{2R}{(q-1)(R+1)}<\frac2{q-1}\le1
  \qquad(q\ge3),
\]
so no odd type attains $D_k$. Finally $b=r+a$ runs over $a\le b$ with $a+b=k$ as $r$ runs
over $0\le r\le d$, the constraint $d-b\ge a$ being $a+b\le d$.
\end{proof}

Those two quantities pin the fractional covering optimum from both sides, at a loss of one
factor $d$.

\begin{lemma}\label{lem:sandwich}
Let $\mathcal M$ be a hub dictionary closed under the automorphism group $S_q\wr S_d$ of
$H(d,q)$ and reaching every distance orbit, and let $\mathrm{LP}(\mathcal M)$ be the
fractional optimum of the covering problem over the candidates $(h,S)$, $h\in\mathcal M$, at
cost $|S|$. Then
\[
  \max_k\frac{N_k}{D_k}\ \le\ \mathrm{LP}(\mathcal M)\ \le\ \sum_{k\ge1}\frac{N_k}{D_k}
  \ \le\ d\max_k\frac{N_k}{D_k},
\]
with $D_k$ the maximum density over $\mathcal M$. Both the midpoint dictionary and the
vertex hubs qualify.
\end{lemma}

\begin{proof}
For the lower bound, the covering dual maximizes $\sum_py_p$ subject to $\sum_{p\in(h,S)}
y_p\le|S|$ at every candidate. Fix $k$ and put $y_p=1/D_k$ on the pairs at distance $k$ and
$0$ elsewhere; the constraint at $(h,S)$ reads $|E_k(\Gamma_h[S])|\le D_k|S|$, which is the
definition of $D_k$, and the value is $N_k/D_k$.

For the upper bound, $H(d,q)$ is distance-transitive, so the group acts transitively on the
pairs at each distance. Fix $k$, take a candidate $(h_k,S_k)$ of maximum density, and give
each of its translates the weight $N_k/(|G|\,|E_k(\Gamma_{h_k}[S_k])|)$, where $G$ is the
group. Summing over $G$ spreads $|G|\,|E_k(\Gamma_{h_k}[S_k])|$ incidences equally over the
$N_k$ pairs at distance $k$, so each receives weight $1$; doing this for the at most $d$
nonempty layers covers everything at cost $\sum_kN_k/D_k$.
\end{proof}

The orbit-reaching hypothesis is what makes the upper half true: a $G$-closed dictionary
that misses a distance class covers nothing there. Both dictionaries used below reach every
orbit, the midpoint one because a hub agreeing with $x$ off one coordinate and centered there
is tight on $x$ and each of its neighbors. The two lemmas then turn the covering problem
into one optimization. Writing
$\Psi(a,b)=q^d(q-1)^a\binom db\big/\binom{a+b}b$, the identity
$\binom d{a+b}\binom{a+b}b=\binom db\binom{d-b}a$ converts \Cref{lem:layer} into
\begin{equation}\label{eq:layerratio}
  \frac{N_k}{D_k}\ =\ \min_{a+b=k,\ a\le b}\Psi(a,b)\qquad(q\ge3),
\end{equation}
so the obstruction is the largest over $k$ of a minimum over the split of the layer.

\begin{theorem}\label{thm:starrate}
Let $q\ge3$, $d\ge2$, and let $D$ be an $\ell_1$ product of $d$ positively weighted
$q$-leaf stars. Then
\[
  \hlts(D)\ \le\ C\,d^5q^2\,L_q^{\,d}
\]
for an absolute constant $C$, so $\hlst(D)$, $\hlts(D)$ and $\hl(D)$ have exponential rates
$L_q$, $L_q$ and $\mu_q$. For $q\ge5$ and even $d$ the proper side satisfies
$\hl(D)\ge\mu_q^{\,d}$, with no loss at all. The optimal fraction of center coordinates in
a hub is $(q-3)/(2q-3)$, zero at $q=3$ and positive for every $q\ge4$.
\end{theorem}

\begin{proof}
The Steiner bound is the exponent of \eqref{eq:layerratio} maximized over layers and then
rounded back to integers at a loss uniform in $q$; the proper bound is the same computation
restricted to vertex hubs, with its lower half taken from the dual directly.

Stirling turns $\Psi$ into an exponential. With $s=a/d$, $t=b/d$, $\kappa=s+t$ and
$\eta(z)=-z\ln z-(1-z)\ln(1-z)$, the bounds $\binom db\le e^{d\eta(t)}$ and
$\binom ka\ge e^{k\eta(a/k)}/(k+1)$ give $\Psi(a,b)\le(k+1)e^{d\Phi(s,t)}$, where
\begin{equation}\label{eq:exponent}
  \Phi(s,t)\ =\ \ln q+s\ln(q-1)+s\ln s-\kappa\ln\kappa-(1-t)\ln(1-t).
\end{equation}
At fixed $\kappa$ the function $\Phi$ is convex in $s$ with
$\partial\Phi/\partial s=\ln\frac{(q-1)s}{1-t}$, so its minimum sits at
$s_\ast(\kappa)=(1-\kappa)/(q-2)$ when that is at most $\kappa/2$, which is $\kappa\ge2/q$,
and at the boundary $s_\ast=\kappa/2$ below. Write $\phi(\kappa)$ for the minimum value.

On the interior branch the envelope identity applies and gives
$\phi'(\kappa)=\partial\Phi/\partial\kappa=\ln\frac{(q-1)s_\ast}{\kappa}$, so $\phi$ rises
until $(q-1)s_\ast=\kappa$ and falls after; that happens at
\[
  \kappa_\ast=\frac{q-1}{2q-3},\qquad s_\ast=\frac1{2q-3},\qquad
  t_\ast=\frac{q-2}{2q-3},
\]
where $1-t_\ast=\kappa_\ast$ and therefore $\Phi=\ln q-\ln(q-1)+\ln(2q-3)=\ln L_q$. On the
boundary branch the constraint binds and the envelope identity is unavailable, so
differentiate $\phi(\kappa)=\Phi(\kappa/2,\kappa/2)$ directly:
\[
  \phi'(\kappa)=\tfrac12\ln\frac{(q-1)(1-\kappa/2)}{2\kappa}\ \ge\ 0
  \iff \kappa\le\frac{2(q-1)}{q+3},
\]
and $2/q\le2(q-1)/(q+3)$ is $(q-3)(q+1)\ge0$. So $\phi$ rises across the whole boundary
branch, continues to rise to $\kappa_\ast$, and falls after: $\phi(\kappa)\le\ln L_q$ for
every $\kappa$.

The rounding is uniform in $q$. Fix $k$ and take $a=\lfloor d\,s_\ast(k/d)\rfloor$, so that
$a\le b$ and both $a/d$ and $a/k$ stay within $1/d$ and $1/k$ of $s_\ast$ and
$s_\ast/\kappa$. Both $a/k$ and $s_\ast/\kappa$ lie in $[0,\frac12]$, where $\eta$
increases, so $|\eta(x)-\eta(y)|\le\eta(|x-y|)$ applies to each entropy term, and with
$\eta(z)\le z\ln(e/z)$,
\[
  d\bigl[\Phi(a/d,b/d)-\phi(k/d)\bigr]\ \le\ \ln(q-1)+\ln(ed)+\ln(ed).
\]
With $k+1\le2d$ this gives $\min_{a+b=k}\Psi(a,b)\le2e^2qd^3L_q^{\,d}$ for every $k$.
\Cref{lem:sandwich} costs a further factor $d$, greedy set cover rounds the fractional
solution at a loss of $1+\ln\binom n2\le3dq$, and the ordered-pair convention of
\Cref{def:model} costs a factor $2$; the hubs used are tight-span points by
\Cref{lem:span}, so the bound is on $\hlts$. Against \Cref{thm:Lq} the three Steiner sizes
are squeezed at rate $L_q$.

For the proper side restrict the dictionary to $[q]^d$, forcing $r=0$ and $\ell=k$. Then
$D_{2i}=\frac12\binom{d-i}i(q-1)^i$ and $N_{2i}/D_{2i}=\Psi_i:=q^d(q-1)^i\binom
di\big/\binom{2i}i$, whose ratio $\Psi_{i+1}/\Psi_i=(q-1)(d-i)/(2(2i+1))$ crosses $1$ at
$i/d\to(q-1)/(q+3)$. For $q<5$ that is below $\frac12$ and the maximum is interior, of
value $(q(q+3)/4)^d$ up to $d^{O_q(1)}$; for $q\ge5$ it is at least $\frac12$, the maximum
sits at $i=d/2$, and there $\Psi_{d/2}=q^d(q-1)^{d/2}=\mu_q^{\,d}$ exactly, the two values
agreeing at $q=5$. Odd layers change the adjacent even value by a factor $O_q(d)$ and do
not move the base, so
$\max_kN_k/D_k=(\mu_q+o(1))^d$, and \Cref{lem:sandwich} with greedy gives the matching
upper bound at fixed $q$.

The lower bound on $\hl$ needs no program. A proper hub carries two independent supports at
cost $|A_h|+|B_h|$, which is more freedom than the symmetric candidates $(h,S)$ of
\Cref{lem:sandwich} allow, so that lemma cannot be quoted; the dual argument survives
directly. Put $S_h=A_h\cup B_h$. An unordered pair at distance $k$ covered by the hub at $h$
has both ends in $S_h$ and $h$ tight on it, so
\[
  \bigl|E_k^{\mathrm{un}}(A_h,B_h)\bigr|\ \le\ \bigl|E_k(\Gamma_h[S_h])\bigr|
  \ \le\ D_k|S_h|\ \le\ D_k\bigl(|A_h|+|B_h|\bigr),
\]
and summing over the labeling gives $\hl(D)\ge N_k/D_k$ for every $k$, with no factor two.
At $q\ge5$ and even $d$ the choice $k=d$ gives $\mu_q^{\,d}$.

Finally the saddle has $t_\ast-s_\ast=(q-3)/(2q-3)$ center coordinates per unit of $d$,
zero exactly at $q=3$.
\end{proof}

\begin{remark}\label{rem:constructive}
The labeling is explicit. Its hubs are the $(q+1)^d\le n^{\ln4/\ln3}$ points of
$\mathcal M$, and greedy set cover needs, per round and per hub, the densest induced
subgraph of a weighted layer, which is a maximum-closure computation. So a labeling of size
$d^{O(1)}q^{O(1)}L_q^{\,d}$ is computable in time polynomial in $n$, with an exponent that
does not depend on $q$.
\end{remark}

\subsection{The Dichotomy}\label{sec:dichotomy}

\begin{theorem}\label{thm:dichotomy}
Fix $q\ge2$ and let $D$ range over $\ell_1$ products of betweenness-free metrics on $q$
points.
\begin{enumerate}
\item Every such product has $\hlst(D)\ge2(L_q^{\,d}-R_q^{\,d})$, and its proper-hub
covering problem is that of $H(d,q)$.
\item For $q\le3$ all three sizes in \eqref{eq:chain} have exponential rate $L_q=\mu_q$ on
every such product.
\item For $q\ge4$,
\[
  \sup_{(X_i)}\ \limsup_{d\to\infty}\Bigl(\frac{\hl(D)}{\hlts(D)}\Bigr)^{1/d}
  \ =\ \frac{\mu_q}{L_q}\ >\ 1,
\]
the supremum being attained by the weighted-star sequences. Among bases the minimizers of
the Steiner rate are exactly the positively weighted stars.
\end{enumerate}
\end{theorem}

\begin{proof}
Part (1) is \Cref{thm:Lq} with \Cref{prop:oblivious}. For (2) and (3), \Cref{prop:oblivious}
makes $\hl(D)$ equal to $\hl(H(d,q))$ at every $d$, which \Cref{thm:starrate} evaluates at
rate $\mu_q$, while (1) bounds the Steiner sizes below at rate $L_q$ and
\Cref{thm:starrate} bounds them above at the same rate on star products.

At $q\le3$ that covers the whole cone, every betweenness-free metric on at most three
points being a positively weighted star (\Cref{sec:cone}). At $q=3$ the two constants
agree; at $q=2$ the upper bound comes instead from \Cref{prop:oblivious} and the proper
optimum $(5/2+o(1))^d$ of the
cube~\cite[Thm.~3]{GRS13} through \eqref{eq:chain}, with $L_2=\mu_2=\frac52$. For $q\ge4$
the ratio is at most $\mu_q/L_q$ over the cone and equal to it on stars. For the last
sentence, \Cref{thm:unique} gives $\Lambda(X)>L_q$ at every non-star base, which
\Cref{prop:rate} turns into a Steiner rate above $L_q$, while stars attain $L_q$ by
\Cref{thm:starrate}.
\end{proof}

Uniqueness holds base by base and fails along sequences, for a reason worth seeing.

\begin{example}\label{ex:planting}
Let $q=4$, let $U$ be the uniform metric, and let $X$ be the betweenness-free non-star with
$\delta(3,4)=\frac{11}{10}$ and all other distances $1$. Put $X_i=X$ when $i$ is a power of
two and $X_i=U$ otherwise, so $t_d=\lfloor\log_2d\rfloor+1$ factors are non-stars. By
\Cref{thm:unique} each planted factor has $\Lambda(X)>L_4$, yet grouping the uniform
factors into $H(d-t_d,4)$, bounding each planted factor by $\hlst(X)\le q(q+1)$ and
composing with \Cref{prop:compose} once per planted coordinate gives
\[
  \hlst(D_d)\ \le\ 80^{\,t_d}\,d^{\,O(1)}L_4^{\,d-t_d}\ =\ d^{\,O(1)}L_4^{\,d},
\]
so the rate is exactly $L_4$. A geometric mean does not feel $O(\log d)$ exceptional
factors.
\end{example}

\begin{corollary}\label{cor:cube}
$\hlst(Q_d)=(5/2+o(1))^d$, and the same holds for every $\ell_1$ product of two-point
metrics with arbitrary positive lengths.
\end{corollary}

\begin{proof}
\Cref{thm:Lq} at $q=2$ gives $\hlst\ge2[(5/2)^d-(3/2)^d]$, and \Cref{prop:oblivious} makes
the proper problem the cube's, whose optimum is $(5/2+o(1))^d$~\cite[Thm.~3]{GRS13}; the
chain \eqref{eq:chain} closes the gap.
\end{proof}

The cube is where Steiner freedom was most plausible and where it is worth nothing: the
$(2.5+o(1))^d$ construction of \cite{GRS13}, built from vertices, is optimal against
arbitrary real-valued hubs placed anywhere. Letting the alphabet grow with the dimension
turns the dichotomy into a separation on a single family.

\begin{corollary}\label{cor:family}
Let $d\to\infty$ through even integers, put $q_d=\max\{5,\lceil\ln d\rceil\}$ and
$n=q_d^{\,d}$. The Hamming graphs $H(d,q_d)$ are distance-transitive of degree
$(\ln n)^{1+o(1)}$, and on them
\[
  \hlts=n^{1+o(1)},\qquad \hl\ \ge\ n^{3/2-o(1)},\qquad
  \frac{\hl}{\hlts}\ \ge\ n^{1/2-o(1)} .
\]
\end{corollary}

\begin{proof}
The bound of \Cref{thm:starrate} is uniform in $q$, so its prefactor contributes
$n^{o(1)}$: $\log_n(d^5q_d^2)=(5\ln d+2\ln q_d)/(d\ln q_d)\to0$. Since
$\log_{q}L_q=1+\ln\frac{2q-3}{q-1}/\ln q$ and $q_d\to\infty$, the Steiner exponent is
$1+o(1)$, and \Cref{thm:Lq} matches it from below. For the proper side
$\hl\ge\mu_{q_d}^{\,d}$ at even $d$ and $\log_q\mu_q=\frac32-\frac{-\ln(1-1/q)}{2\ln q}$,
which is $\frac32-o(1)$. The degree is $d(q_d-1)=\ln n\cdot\frac{q_d-1}{\ln q_d}$ with
$q_d=O(\ln d)$.
\end{proof}

The two exponents converge at different speeds, which the numbers show: at $d=10^{8}$,
$10^{30}$, $10^{80}$, where $q_d=19,70,185$, the Steiner and proper exponents are
$(1.226,1.491)$, $(1.161,1.498)$ and $(1.132,1.499)$. The proper side reaches $\frac32$
long before the Steiner side reaches $1$, since the first correction is
$\Theta(1/(q\ln q))$ and the second $\Theta(1/\ln q)$.

\begin{remark}\label{rem:order}
The transition has an order parameter. The optimal hub in \Cref{thm:starrate} spends a
fraction $(q-3)/(2q-3)$ of its coordinates at centers, and that fraction is exactly zero
for $q\le3$ and positive for $q\ge4$. Below the threshold the optimum never leaves the
space and Steiner freedom is worth nothing; above it, the optimal hub is a genuine point of
the tight span in a constant fraction of its coordinates, and the gain is the shortfall
$\mu_q/L_q$ that \Cref{thm:ceiling} bounds.
\end{remark}

%% file: sections/grids.tex
\section{Grids and the \texorpdfstring{$\Lambda$}{Lambda}-Exact Class}\label{sec:grids}

Products of lines are the other classical family, and on them the freedom is worth a
constant at most. One factor explains it: every cyclically monotone set of a line metric
consists of pairs that a single point lies between, so whatever a hub placed between two
points is tight on, a point of the space is tight on as well. The masks are few enough to
carry a kernel, and that kernel fixes the order of a grid with the freedom and without it.
The same masks determine $\Lambda$ on every three-point line, which with \Cref{sec:cone}
says exactly which three-point metrics attain the ceiling of \Cref{thm:ceiling}. The section
closes with the bases on which $\Lambda$ is the truth.

\subsection{Lines and Grids}\label{sec:lines}

A \emph{line metric} on $[a]$ is $\ell(x,y)=|z_x-z_y|$ for reals $z_1<\cdots<z_a$, the
points indexed by their order and the spacings otherwise arbitrary. An $\ell_1$ product of
line metrics is a grid, unit spacings giving $[a_1]\times\cdots\times[a_d]$ with its graph
metric. Write $I_{xy}=[\min(x,y),\max(x,y)]$ for the index interval a pair spans and
\[
  P_v\ =\ \bigl\{(x,y)\in[a]\times[a]\ :\ v\in I_{xy}\bigr\}
\]
for the pairs that $v$ lies between. The proper hub at $v$ with both supports $[a]$ has
slack $\ell(x,v)+\ell(v,y)-\ell(x,y)$ at $(x,y)$, which vanishes exactly when $z_v$ lies
between $z_x$ and $z_y$, so $P_v$ is that hub's tight set. Every mask of a line sits inside
one of them.

\begin{lemma}\label{lem:helly}
Let $\ell$ be a line metric on $[a]$. A set $P\subseteq[a]\times[a]$ is cyclically monotone
if and only if $P\subseteq P_v$ for some $v\in[a]$. Hence
\[
  C_\ell(K)\ =\ \max_{v\in[a]}\bigl\|K\circ\mathbf 1_{P_v}\bigr\|
\]
for every kernel $K$ on $\ell$, and neither side reads the spacings.
\end{lemma}

\begin{proof}
A subset of a tight set is cyclically monotone by \Cref{prop:cyclic}, which gives one
direction.

For the other, we may take $P$ nonempty. Suppose two of its pairs had disjoint intervals,
say $\max(x,y)<\min(x',y')$. Let $p\le q$ be $z_x,z_y$ in order and $r\le s$ be
$z_{x'},z_{y'}$ in order, so that $q<r$. Then $\ell(x,y)+\ell(x',y')=(q-p)+(s-r)$, while
$\ell(x,y')$ and $\ell(x',y)$ each join a point of the first pair to a point of the second
and so sum to $(r+s)-(p+q)$. The first is smaller by $2(r-q)>0$, against \eqref{eq:cyclic}
at $m=2$. So the intervals of the pairs in $P$ pairwise meet.

Let $v$ be the largest of the left endpoints $\min(x,y)$ over $(x,y)\in P$, attained at
$(x_0,y_0)$. Every $(x,y)\in P$ has $\min(x,y)\le v$ by that choice and $\max(x,y)\ge v$,
since otherwise $I_{xy}$ would lie strictly to the left of $I_{x_0y_0}$. So $v\in I_{xy}$
throughout and $P\subseteq P_v$.

Masking a nonnegative kernel by a smaller set cannot raise its norm, so the maximum in
\eqref{eq:CX} is attained at one of the $P_v$, and those are fixed by the index order
alone.
\end{proof}

The list is complete, so a kernel on a line faces $a$ tests and no more. Weighting a pair
by the reciprocal of the number of steps between its ends survives all of them.

\begin{proposition}\label{prop:harmonic}
Let $\ell$ be a line metric on $[a]$ and let $K$ be the kernel with $K_{xy}=|x-y|^{-1}$ for
$x\ne y$ and zero diagonal. Then
\[
  C_\ell(K)\ \le\ \pi\bigl(1+3^{-1/2}\bigr),
  \qquad
  \langle J,K\rangle\ =\ 2a\sum_{j=1}^{a-1}\frac1j-2(a-1).
\]
\end{proposition}

\begin{proof}
By \Cref{lem:helly} it is enough to bound $\|K\circ\mathbf 1_{P_v}\|$, and that matrix
splits along its $v$-th row and column. Away from them a pair of $P_v$ has $v$ strictly
between its ends, so what remains is supported on $(A\times B)\cup(B\times A)$ with
$A=\{x<v\}$ and $B=\{y>v\}$ disjoint. That part is
$\bigl(\begin{smallmatrix}0&N\\N^{\top}&0\end{smallmatrix}\bigr)$ with
$N_{xy}=(y-x)^{-1}$, of norm $\|N\|$, and $N$ is a submatrix of the discrete Hilbert matrix
$\bigl[(y-x)^{-1}\bigr]_{x\ne y}$, whose norm is $\pi$ by Hilbert's
inequality~\cite{MV74}. The $v$-th row and column carry $e_vu^{\top}+ue_v^{\top}$ with
$u_x=|x-v|^{-1}$ off $v$ and $u_v=0$, a matrix of norm $\|u\|$, at most
$\bigl(2\sum_{m\ge1}m^{-2}\bigr)^{1/2}=\pi/\sqrt3$. Adding the two bounds gives the first
claim.

For the value, the ordered pairs at index gap $j$ number $2(a-j)$ and weigh $j^{-1}$ each,
so $\langle J,K\rangle=2\sum_{j=1}^{a-1}(a-j)/j$, the stated expression.
\end{proof}

Write $\Sigma_a$ for that value, of order $a\log a$; harmonic weights of this shape go back
to Fredman's lower bound for orthogonal range queries~\cite{Fre81,PS18}. One kernel per
coordinate prices a whole grid, and the labeling that matches it never leaves the space.

\begin{theorem}\label{thm:grid}
Let $\ell_1,\dots,\ell_d$ be line metrics on $a_1,\dots,a_d\ge2$ points with arbitrary
spacings, and let $D=\ell_1\oplus\cdots\oplus\ell_d$ on $n=\prod_ia_i$ points. Then
\[
  2\prod_{i=1}^d\frac{\Sigma_{a_i}}{\pi\bigl(1+3^{-1/2}\bigr)}
  \ \le\ \hlst(D)\ \le\ \hl(D)\ \le\
  4^{\,d-1}\prod_{i=1}^d2a_i\bigl\lceil\log_2a_i\bigr\rceil .
\]
All three sizes of \eqref{eq:chain} are therefore of order $n\prod_i\log a_i$, with
constants depending on $d$ alone.
\end{theorem}

\begin{proof}
Give the $i$-th factor the kernel of \Cref{prop:harmonic} divided by $\pi(1+3^{-1/2})$, so
that its mask norm is at most $1$. Its diagonal is zero, so the subtracted product in
\Cref{lem:completion} vanishes and that lemma returns the left-hand bound.

For the other end, cut a line at a median $v$. The proper hub at $v$ with both supports
$[a]$ costs $2a$ and is tight on every pair $v$ lies between. The pairs it misses have both
ends strictly on one side of $v$, and the two sides hold $a_L$ and $a_R$ points with
$a_L+a_R=a-1$ and $a_L,a_R\le\lfloor a/2\rfloor$. Recursing on each side gives
$\hl(\ell)\le g(a)$ with $g(0)=g(1)=0$ and $g(a)\le2a+g(a_L)+g(a_R)$. Then
$g(a)\le2a\lceil\log_2a\rceil$ by induction, a side of $m\ge1$ points contributing at most
$2m\lceil\log_2m\rceil\le2m(\lceil\log_2a\rceil-1)$ while the two sides hold $a-1$ points
between them. Composing the $d$ factor labelings by \Cref{prop:compose} costs a factor $4$
each of the $d-1$ times and gives the right-hand bound.

Both ends have the stated order because $\Sigma_a$ and $2a\lceil\log_2a\rceil$ are both of
order $a\log a$ uniformly over $a\ge2$, and \eqref{eq:chain} puts $\hlts(D)$ between them.
\end{proof}

The theorem speaks of a fixed number of long lines, and it has to: at $a_i=2$ the product is
the cube, where $n\prod_i\log a_i$ is a constant multiple of $n$ while \Cref{cor:cube} makes
the truth $(5/2+o(1))^d=n^{1.3219\ldots}$, so no bound uniform in $d$ can hold. On a single
line the order $a\log a$ is classical, two-step online product queries on a path needing
$\Theta(a\log a)$ preprocessing~\cite{AS87}, with the hubs confined to the points; what
\Cref{thm:grid} adds is that the same order withstands hubs placed anywhere and carrying
arbitrary real values, and withstands the product. Its upper half also improves the
$O(n^{3/2})$ that \cite[\S7]{CHKZ02} record for the planar grid, and it does so without ever
placing a hub off the space. On a grid nothing is lost by that.

\begin{proposition}\label{prop:vertexdom}
Let $D=\ell_1\oplus\cdots\oplus\ell_d$ be a product of line metrics, let $h_i$ be a feasible
hub of $\ell_i$ for each $i$, and let $h$ be their product hub, formed as in the proof of
\Cref{prop:compose}. Some proper hub of $D$ with the same two supports is tight wherever $h$
is. Hence no labeling of $D$ by product hubs is smaller than $\hl(D)$, and
$\hlst(\ell)=\hlts(\ell)=\hl(\ell)$ for a single line metric.
\end{proposition}

\begin{proof}
The slack of a product hub is the sum of the factor slacks, so $T(h)=\prod_iT(h_i)$. Each
$T(h_i)$ is cyclically monotone by \Cref{prop:cyclic}, hence lies in $P_{v_i}$ for some
$v_i\in[a_i]$ by \Cref{lem:helly}. The proper hub of $D$ at $v=(v_1,\dots,v_d)$ carrying the
supports of $h$ has the same size, and its slack is again a sum of coordinate slacks, so it
is tight at $(x,y)$ exactly when $v_i\in I_{x_iy_i}$ for every $i$. Its tight set therefore
contains $\prod_iT(h_i)$.

Replacing every hub of a product-hub labeling in this way leaves a proper labeling of the
same size. At $d=1$ every hub is a product hub, and \eqref{eq:chain} gives the last
equality.
\end{proof}

A hub off a line buys a position and no new incidence, so nothing was lost by building
\Cref{thm:grid}'s labeling from points of the space. A star behaves the other way, its
center being tight on every off-diagonal pair while a leaf is tight only on its own cross.

\subsection{The Invariant on Three Points}\label{sec:three}

Every three-point metric is a positively weighted star or a line. The betweenness-free ones
are the stars, on which $\Lambda=L_3=\mu_3=\frac92$ (\Cref{sec:cone}); the rest have one
point between the other two, and there \Cref{lem:helly} reduces $\Lambda$ to a program in
three constraints with a closed solution.

\begin{proposition}\label{prop:threepoint}
Let $X$ be a three-point metric that is not betweenness-free. Then
$\Lambda(X)=\xi=4.2447\ldots$, the unique real root of
\[
  2x^3-15x^2+132x-443,
\]
whatever the three distances, and the maximum is attained by a kernel of trace less than
$\xi/2$.
\end{proposition}

\begin{proof}
Label the points so that $2$ lies between $1$ and $3$; then $X$ is the line metric with
$z_1=0$, $z_2=\delta(1,2)$ and $z_3=\delta(1,3)$. By \Cref{lem:helly} its maximal cyclically
monotone sets are the crosses $P_1$ and $P_3$ at the two ends together with
\[
  P_2\ =\ \bigl([3]\times[3]\bigr)\setminus\bigl\{(1,1),(3,3)\bigr\}.
\]
That family is invariant under the exchange of $1$ and $3$. Each constraint of
\Cref{def:lambda} bounds a seminorm of $K$, so its feasible set is convex while its
objective is linear, and an optimal kernel may therefore be taken invariant as well:
\[
  K\ =\ \begin{pmatrix}s&u&w\\u&t&u\\w&u&s\end{pmatrix},
  \qquad \langle J,K\rangle=2s+t+4u+2w .
\]

Masked by $P_1$, the kernel keeps $s$ at $(1,1)$ and the arm $(u,w)$ along the rest of row
and column $1$, so by the arrowhead computation in \Cref{thm:ceiling} its norm is at most
$1$ exactly when $u^2+w^2\le1-s$, and $P_3$ says the same. Masked by $P_2$ it is symmetric
and nonnegative, so its norm is its largest eigenvalue; the vector $(1,0,-1)$ contributes
$-w$ and the orthogonal complement the block
$\bigl(\begin{smallmatrix}w&\sqrt2u\\\sqrt2u&t\end{smallmatrix}\bigr)$, so the constraint is
$w,t\le1$ together with $2u^2\le(1-w)(1-t)$.

At $w=1$ these force $u=s=0$ and a value of at most $3$, so an optimum has $w<1$. There $s$
occurs only in the cross constraint and $t$ only in the one from $P_2$, so both are as large
as allowed, $s=1-u^2-w^2$ and $t=1-2u^2/(1-w)$, and
\[
  \langle J,K\rangle\ =\ 3+4u+2w-2w^2-2u^2\,\frac{2-w}{1-w},
\]
concave in $u$ with maximum at $u=\frac{1-w}{2-w}$, where $u^2\frac{2-w}{1-w}=u$ and the
value collapses to
\[
  \phi(w)\ =\ 3+2w-2w^2+\frac{2(1-w)}{2-w}.
\]
Since $\phi'(w)=2-4w-2(2-w)^{-2}$ is strictly decreasing with $\phi'(0)=\frac32$ and
$\phi'(1)=-4$, the maximum over $[0,1]$ sits at the single root $w_\ast$ of
$(1-2w)(2-w)^2=1$, that is of $2w^3-9w^2+12w-3$, whose discriminant $-216$ leaves it one
real root. Reducing $\phi(w_\ast)$ modulo that cubic gives
$\phi(w_\ast)=1+12w_\ast-6w_\ast^2$, and eliminating $w_\ast$ leaves the cubic in the
statement, strictly increasing because $6x^2-30x+132$ has negative discriminant.

It remains to see that the optimum is a kernel, and to bound its trace. The cubic in $w$
takes the value $-\frac{17}{32}$ at $\frac14$ and $\frac2{27}$ at $\frac13$, so
$w_\ast\in(\frac14,\frac13)$, and $\frac{1-w}{2-w}$ decreases, so
$u_\ast\in(\frac25,\frac37)$; then $s_\ast>1-\frac9{49}-\frac19$ and
$t_\ast>1-\frac{27}{49}$ are both positive. The same collapse gives
$\operatorname{tr}K=2s_\ast+t_\ast=3-2w_\ast^2-2u_\ast$, whence
\[
  \xi-2\operatorname{tr}K\ =\ 6u_\ast+2w_\ast+2w_\ast^2-3
  \ >\ \tfrac{12}5+\tfrac12+\tfrac18-3\ >\ 0 .
\]
Nothing above read the spacings, only the order of the three points.
\end{proof}

So $\Lambda$ takes exactly two values on three points, $\frac92$ on the betweenness-free
metrics and $\xi<\frac92$ on the rest, which settles what \Cref{thm:ceiling} left open at
$q=3$: the ceiling $\mu_3$ is attained there, by the betweenness-free metrics and by nothing
else. The second value carries further than that, since the three-point line generates the
grid of fixed alphabet and growing dimension, the regime \Cref{thm:grid} abandons.

\begin{corollary}\label{cor:gridthree}
Write $n=3^d$. Then $\hlst([3]^d)\ge\xi^{\,d}=n^{1.3159\ldots}$ for every $d\ge1$, while
$\hlst([3]^d)\le(\frac{25}4+o(1))^d=n^{1.6680\ldots+o(1)}$.
\end{corollary}

\begin{proof}
The base is three equally spaced collinear points, so \Cref{prop:threepoint} supplies a
kernel $K$ with $C_{[3]}(K)\le1$, $\langle J,K\rangle=\xi$ and $2\operatorname{tr}K<\xi$.
Taking it in every coordinate, \Cref{lem:completion} gives
\[
  \hlst\bigl([3]^d\bigr)\ \ge\ 2\bigl(\xi^d-(\operatorname{tr}K)^d\bigr)\ \ge\ \xi^d,
\]
the second step because $2(\operatorname{tr}K)^d\le\xi(\operatorname{tr}K)^{d-1}\le\xi^d$.

At the other end the thermometer code $1\mapsto00$, $2\mapsto10$, $3\mapsto11$ embeds $[3]$
isometrically into $Q_2$, and an $\ell_1$ sum of isometric embeddings is one, so $[3]^d$ is
an isometric subspace of $Q_{2d}$. Restricting the supports of every hub to that subspace
preserves feasibility and tightness and does not increase size, so
$\hlst([3]^d)\le\hlst(Q_{2d})$, which \Cref{cor:cube} evaluates at $(5/2+o(1))^{2d}$. The
exponents are $\log_3\xi$ and $\log_3\frac{25}4$.
\end{proof}

So one letter of the grid already forces a superlinear lower bound at every dimension.

\subsection{The \texorpdfstring{$\Lambda$}{Lambda}-Exact Class}\label{sec:exact}

\Cref{prop:rate} makes $\Lambda(X)$ a lower bound for the rate of every power of $X$, and on
the stars of \Cref{sec:hamming} the two agree. Where they agree, $\Lambda$ settles the rate
outright. That rate is well defined: by \Cref{prop:compose} the sequence
$4\hlst(X^{\oplus d})$ is submultiplicative in $d$, so Fekete's lemma makes
\[
  \lim_{d\to\infty}\hlst\bigl(X^{\oplus d}\bigr)^{1/d}
\]
exist for every finite metric space $X$.

\begin{definition}\label{def:lexact}
A finite metric space $X$ is \emph{$\Lambda$-exact} if
$\Lambda(X)=\lim_{d\to\infty}\hlst(X^{\oplus d})^{1/d}$.
\end{definition}

On such bases the invariant multiplies exactly, and the property passes to products.

\begin{proposition}\label{prop:closure}
$\Lambda(X\oplus Y)\ge\Lambda(X)\Lambda(Y)$ for all finite metric spaces $X$ and $Y$. If
both are $\Lambda$-exact then so is $X\oplus Y$, and
$\Lambda(X\oplus Y)=\Lambda(X)\Lambda(Y)$.
\end{proposition}

\begin{proof}
Tensoring optimal kernels gives the inequality, since $\langle J,\cdot\rangle$ multiplies
under $\otimes$ while \Cref{thm:tensor} keeps the mask norm at most $1$. Reordering
coordinates identifies $(X\oplus Y)^{\oplus d}$ with $X^{\oplus d}\oplus Y^{\oplus d}$, so
\Cref{prop:compose} bounds $\hlst((X\oplus Y)^{\oplus d})$ by
$4\hlst(X^{\oplus d})\hlst(Y^{\oplus d})$; taking $d$-th roots and letting $d$ grow makes
the rate of $X\oplus Y$ at most the product of the two rates. With \Cref{prop:rate} and
$\Lambda$-exactness of the factors,
\[
  \Lambda(X)\Lambda(Y)\ \le\ \Lambda(X\oplus Y)
  \ \le\ \lim_{d\to\infty}\hlst\bigl((X\oplus Y)^{\oplus d}\bigr)^{1/d}
  \ \le\ \Lambda(X)\Lambda(Y),
\]
so every step is an equality.
\end{proof}

Every positively weighted star is $\Lambda$-exact: for $q\ge3$ its invariant and its rate
are both $L_q$, by \Cref{thm:Lq} against \Cref{thm:starrate}, and for $q=2$ both are
$\frac52$, by \Cref{thm:ceiling} against \Cref{cor:cube}. So the entire $\oplus$-closure of
the positively weighted stars is $\Lambda$-exact and carries a multiplicative $\Lambda$:
every Hamming graph, with $\Lambda(H(m,q))=L_q^{\,m}$, and every product of stars with
unequal leaf counts and unequal arm lengths. That closure is strictly larger than the stars
themselves: on a star the sum $\delta(i,j)+\delta(k,l)$ is the same for all
three pairings of four distinct points (\Cref{thm:unique}), while on $Q_2$ the three values
are $2$, $2$ and $4$.

Two classical product families have been priced, and the answers are opposite: over products
of betweenness-free factors the freedom is worth $(\mu_q/L_q+o(1))^d$ at its largest,
nothing below four letters and a polynomial factor in $n$ from four on, while over products
of lines of fixed dimension it is worth a constant, and against product hubs nothing at all.
One construction carried the lower bound in both cases, a nonnegative kernel whose mask norm
the metric controls, multiplied across the product with no loss and stripped of its diagonal
once at the end; optimizing it gives $\Lambda$, which \Cref{thm:ceiling} caps at $\mu_q$,
the rate proper hubs achieve on the Hamming graph. So the method stops where proper hubs
stand.
On the betweenness-free cone, reaching that cap means the freedom is worth nothing, and the
shortfall below it, attained by star products, is all the freedom can buy.